%% file: paper.tex
\PassOptionsToPackage{table}{xcolor}
\documentclass[sigconf]{acmart}
\pdfoutput=1

\input{package-setup}
\input{listing-setup}

\copyrightyear{2021} 
\acmYear{2021} 
\setcopyright{acmlicensed}\acmConference[ACSAC '21]{Annual Computer Security Applications Conference}{December 6--10, 2021}{Virtual Event, USA}
\acmBooktitle{Annual Computer Security Applications Conference (ACSAC '21), December 6--10, 2021, Virtual Event, USA}
\acmPrice{15.00}
\acmDOI{10.1145/3485832.3485884}
\acmISBN{978-1-4503-8579-4/21/12}

\begin{document}

\title[The Emperor's New Autofill Framework: A Security Analysis of Autofill on iOS and Android]{The Emperor's New Autofill Framework:\\A Security Analysis of Autofill on iOS and Android}

\author{Sean Oesch}
\email{toesch1@vols.utk.edu}
\affiliation{
	\institution{The University of Tennessee}
	\city{Knoxville}
	\state{Tennessee}
	\country{USA}
}

\author{Anuj Gautam}
\email{anujgtm123@vols.utk.edu}
\affiliation{
	\institution{The University of Tennessee}
	\city{Knoxville}
	\state{Tennessee}
	\country{USA}
}

\author{Scott Ruoti}
\email{ruoti@utk.edu}
\affiliation{
	\institution{The University of Tennessee}
	\city{Knoxville}
	\state{Tennessee}
	\country{USA}
}

\begin{abstract}
\input{abstract}
\end{abstract}



\maketitle

\input{introduction}
\input{background}

\input{methodology}
\input{browser}
\input{app}
\input{webview}

\input{rw}
\input{discussion}
\input{conclusion}

\section*{Responsible Disclosure}
We have informed the developers of the frameworks and password managers we evaluated of our results and have notified both Apple and Google of the WebView-based attacks we discovered.

\section*{Research Artifacts}
The generated data, scripts used to analyze that data, and all analysis artifacts will be available for download at \url{[redacted]}.


\bibliographystyle{plain}
\bibliography{paper}

\appendix
\input{appendix-downloads}
\input{appendix-results}
\input{appendix-mapping}

\end{document}

%% file: package-setup.tex
\usepackage{balance}
\usepackage{enumitem}
\usepackage{graphicx}
\usepackage{microtype}
\usepackage{subcaption} %

\microtypecontext{spacing=nonfrench}
\emergencystretch=\maxdimen
\usepackage{tcolorbox} 
\usepackage{xspace}

\def\findingboxmargin{3pt}
\newtcolorbox{findingbox}{colback=blue!5!white,top=\findingboxmargin,bottom=\findingboxmargin,left=\findingboxmargin,right=\findingboxmargin}

\newcounter{findingcounter}
\newcommand*{\finding}[1]{\begin{findingbox}{\textbf{Finding \#\thefindingcounter}:}\stepcounter{findingcounter}\xspace#1\end{findingbox}}

\usepackage{adjustbox} %
\usepackage{array}
\usepackage{hhline}
\usepackage{multirow}

\newcommand*{\setuptable}{
	\setlength{\arrayrulewidth}{0.1em}
	\centering
}

\newcommand*\rot{\rotatebox{90}}
\newcommand*{\headrow}[1]{\rot{#1}}
\newcommand*{\headcol}[2]{\multirow{#1}{*}{\rot{#2}}}

\newcommand*{\prtline}{}

\usepackage{wasysym} %
\newcommand*{\full}{\CIRCLE}
\newcommand*{\prt}{\LEFTcircle}
\newcommand*{\none}{\Circle}
\newcommand*{\edit}{\ding{46}}
\newcommand*{\na}{}

\usepackage{pifont} %
\renewcommand*{\checked}{\ding{52}}

\newcommand*{\prop}[1]{P#1}

%% file: listing-setup.tex
\usepackage{listings}

\definecolor{backgroundColor}{HTML}{fafafa}
\definecolor{keywordColor}{HTML}{d73a49}
\definecolor{stringColor}{HTML}{032f62}
\definecolor{commentColor}{HTML}{606770}
\definecolor{numberColor}{HTML}{D33682}
\colorlet{punctuationColor}{red!60!black}
\definecolor{scopingColor}{RGB}{20,105,176}

\lstdefinelanguage{JavaScript}{
	morekeywords={typeof, new, true, false, catch, let, function, func, return, null, catch, switch, var, if, in, while, do, else, case, break},
	morecomment=[s]{/*}{*/},
	morecomment=[l]//,
	morestring=[b]",
	morestring=[b]'
}

\lstdefinelanguage{json}{
	morestring=[b]",
	morestring=[b]',
}

\lstdefinelanguage{swift}
{
  morekeywords={
    func,if,then,else,for,in,while,do,switch,case,default,where,break,continue,fallthrough,return,
    typealias,struct,class,enum,protocol,var,func,let,get,set,willSet,didSet,inout,init,deinit,extension,
    subscript,prefix,operator,infix,postfix,precedence,associativity,left,right,none,convenience,dynamic,
    final,lazy,mutating,nonmutating,optional,override,required,static,unowned,safe,weak,internal,
    private,public,is,as,self,unsafe,dynamicType,true,false,nil,Type,Protocol,
  },
  morecomment=[l]{//}, 
  morecomment=[s]{/*}{*/}, 
  morestring=[b]" 
}

%% file: abstract.tex

Password managers help users more effectively manage their passwords, encouraging them to adopt stronger passwords across their many accounts.
In contrast to desktop systems where password managers receive no system-level support, mobile operating systems provide autofill frameworks designed to integrate with password managers to provide secure and usable autofill for browsers and other apps installed on mobile devices.
In this paper, we evaluate mobile autofill frameworks on iOS and Android, examining whether they achieve substantive benefits over the ad-hoc desktop environment or become a problematic single point of failure.
Our results find that while the frameworks address several common issues, they also enforce insecure behavior and fail to provide password managers sufficient information to override the frameworks' insecure behavior, resulting in mobile managers being less secure than their desktop counterparts overall.
We also demonstrate how these frameworks act as a confused deputy in manager-assisted credential phishing attacks.
Our results demonstrate the need for significant improvements to mobile autofill frameworks.
We conclude the paper with recommendations for the design and implementation of secure autofill frameworks.

%% file: introduction.tex

\section{Introduction}
The cognitive burden of remembering many strong, unique passwords leads users to create easily guessed passwords~\cite{dell2010password,riley2006password} and to reuse passwords~\cite{das2014tangled,wang2018end,pearman2017let,florencio2007large}.
These insecure behaviors make targeted attacks easier and lead to large-scale account compromise when data breaches occur.
While other authentication schemes have been proposed, passwords remain dominant~\cite{bonneau2012quest,bonneau2012science}.

Password managers offer a pathway to help users more effectively manage their passwords, assisting users to create strong passwords, store those passwords, and finally fill those passwords into login forms (i.e., password autofill), significantly reducing the cognitive burden of using strong, unique passwords~\cite{zhang2007cantina,lyastani2018better}.
On the other hand, if implemented incorrectly, password managers can become a single point of failure, putting all a user's credentials at risk~\cite{oesch2020that}.
To secure autofill, password managers must only fill credentials when: (\textbf{\prop{1}}) the user has explicitly authorized the fill operation~\cite{li2014emperor,silver2014password}, (\textbf{\prop{2}}) the credential is mapped to the web domain or app to be filled~\cite{aonzo2018phishing,oesch2020that}, and (\textbf{\prop{3}}) the filled credential will only be accessible to the mapped app or web domain.~\cite{stock2014protecting}.

On desktop environments, password managers are primarily implemented as ad-hoc browser extensions---i.e., the extension individually implements all aspects of the autofill process without support from OS or browser autofill frameworks.
While some desktop password managers correctly achieve \prop{1} and \prop{2}~\cite{oesch2020that}, many have incorrect implementations that allow attackers to steal or phish users' credentials~\cite{li2014emperor,silver2014password,stock2014protecting,oesch2020that}, and none can fully implement \prop{3} due to technical limitations of browser extension APIs~\cite{stock2014protecting,oesch2020that}.

In contrast to the desktop environment, mobile operating systems provide system-wide autofill frameworks that attempt to standardize and secure the autofill process.
Critically, these frameworks have the potential to enforce correct handling of \prop{1}--\prop{3} for all mobile password managers.
Additionally, these frameworks provide support for autofill within apps, which is largely unavailable on desktops.

\emph{In this paper, we conduct the first evaluation of the mobile autofill frameworks, examining whether they achieve substantive benefits over the ad-hoc desktop environment or instead become a problematic single point of failure.}
In this evaluation, we consider all such frameworks: iOS's app extensions, iOS's Password AutoFill, and Android's autofill service.
Positively, our evaluation finds that all frameworks correctly require user interaction before autofilling credentials (\prop{1}), a marked improvement over the mixed support on desktop.
In contrast, we find that framework support for \prop{2} and \prop{3} is severely limited.

Within browsers, we find that the frameworks do not correctly validate the authenticity of the webpage nor ensure that filled credentials will be sent to the appropriate domain upon form submission.
The frameworks also provide minimal information about the webpage to the password managers, preventing them from making appropriate security checks.
Even with improvements for \prop{1}, this leaves mobile password managers less secure than their desktop counterparts.

Within apps, autofill behavior differs based on the type of interface being filled: (1) native UI elements, (2) WebView controls, and (3) custom-drawn UI elements, of which mobile autofill frameworks support the first two.
For native elements, we find that iOS Password AutoFill provides a secure and robust binding between credentials and apps, achieving \prop{2} and \prop{3}.
In contrast, the Android autofill service provides no such binding, leaving the mapping of credentials and apps to the individual password managers, with nearly all such mappings being insecure (breaking \prop{2}).
Even worse, iOS app extensions fail to provide a secure mapping mechanism and prevent managers from implementing their own mappings, allowing any credential to be autofilled into any app.

For WebView controls, credentials should only be autofilled if they match the webpage's domain within the WebView control, regardless of which credentials are mapped to the app.
Such behavior is enforced by iOS Password Autofill (achieving \prop{2}), whereas both iOS app extensions and the Android autofill service leave the mapping decision to the individual password managers, with only some managers implementing the mapping correctly.
Managers that do not correctly implement this mapping instead autofill the app's mapped credentials into the webpage, allowing malicious or compromised webpages displayed in benign apps to phish the app's mapped credential (breaking \prop{3}).
Even in the frameworks and managers that do enforce a secure mapping, we identify a limitation in the design of WebView controls on Android and iOS that allows a malicious app to host benign webpages within a (potentially invisible) WebView and steal filled credentials, enabling the surreptitious phishing of all the user's credentials (also breaking \prop{3}).

Critically, in both phishing attacks, the password manager acts as a confused deputy, displaying the autofill dialog and suggesting that the user fills the credential being targeted by the attack.
This behavior is highly problematic, as, in all other contexts, the autofill dialog is an indication that phishing is not occurring and is designed to give users confidence in filling their credentials.
As such, users are unlikely to look carefully at the autofill dialog, increasing the probability that their credentials will be successfully stolen.

Overall, our results show significant security issues with mobile autofill frameworks, limiting both the utility and usability (as users need to be ever vigilant) of mobile password managers.
This situation is especially problematic as password managers are being promoted frequently by the news media and security experts.
Still, all is not lost, and with improvements, these frameworks could become the most secure way to implement autofill on both mobile and desktop.
We conclude this paper by providing recommendations for improving autofill frameworks and detail how the design of WebView controls could be changed to address the phishing attacks identified in this paper.

%% file: background.tex

\section{Background}
\label{sec:background}

Password managers serve to help users (a) create random, unique credentials for each service they authenticate to, (b) store the user's credentials (both generated and user-entered), and (c) fill those credentials for the user.
%

On desktop environments, password managers are implemented as ad-hoc browser extensions (i.e., the browser provides no first-party APIs supporting password management).
In contrast, on mobile, there are first-party frameworks that assist managers in conducting the autofill process.
This first-party support allows for autofill both within browsers (see \S\ref{sec:browser}) and within apps, something largely not possible with desktop managers.

For apps, there are three types of interfaces where autofill would be applicable: (1) native UI elements (i.e., OS-provided widgets), (2) using custom UI elements drawn and managed by the app, and (3) within a webpage displayed in a WebView hosted by the app.
The security of autofill for native UI elements is discussed in \S\ref{sec:app}, while the security of autofill in WebView controls is discussed in \ref{sec:webview}.
Mobile autofill frameworks do not support custom rendered UI elements.

\subsection{Secure Autofill}
\label{sec:secure-autofill}


By examining past research~\cite{li2014emperor,silver2014password,stock2014protecting,aonzo2018phishing,oesch2020that}, we have synthesized three properties that need to be guaranteed by password managers in order for the autofill operation to be secure:

	\textbf{\prop{1}---User authorization.}
	Requiring user authorization before filling credentials helps reduce the attack surface by limiting how often credentials are filled and thus vulnerable to theft~\cite{li2014emperor,silver2014password,oesch2020that}.
	Without interaction, credentials would be automatically filled into any login form, leaving them vulnerable to theft if an attacker can control the page's contents (for example, XSS vulnerabilities remain common~\cite{xssAttacks}, supply chain attacks are increasingly regular~\cite{supplyChain}, and network attacks are feasible when users connect to public WiFi access points~\cite{wifiSecurity}).
	
	Note, it should not be assumed that users will carefully examine these dialogs, as they likely will become habituated to clicking through them~\cite{dhamija2006phishing,felt2011phishing}.
	Instead, requiring interaction is primarily intended to prevent the surreptitious entry and subsequent theft of credentials---for example, Firefox's built-in password manager fails to require user interaction and due to other flaws in its implementation allows an adversary to silently steal all of the user's credentials~\cite{oesch2020that}; such an attack would be difficult if not impossible to conduct if user interaction was required as the attack would trigger thousands of autofill requests, alerting the user that something was wrong~\cite{oesch2020that}.
	Similarly, requiring interaction can also help prevent attacks when the user is not trying to authenticate, in which case they are more likely to click out of the autofill interface to resume their usage of the webpage or app, thus preventing potential credential theft.
	
	\textbf{\prop{2}---Secure credential to destination mapping.}
	Managers need to be able to map credentials to the webpage and apps they should be filled in, preventing other webpages and apps from accessing those credentials~\cite{aonzo2018phishing}.
	Such a mapping prevents malicious webpages and apps from stealing credentials intended for other webpages and apps---i.e., phishing attacks.
	This mapping is commonly done by associating credentials with domains and then associating those domains with apps.
	\textit{When properly implemented, the autofill interface then becomes a sign to users that they are not being phished and can safely enter the suggested credentials on the current site or app.}
	
	\textbf{\prop{3}---Credentials are only accessible to mapped destinations.}
	Managers should ensure that after credentials are filled, they will only be accessible to mapped destinations~\cite{stock2014protecting,oesch2020that}.
	For webpages, that means that the credentials will only be sent to a server on the same domain and that they will not be accessible to malicious JavaScript running on the page that may try to exfiltrate the credentials to different domains~\cite{stock2014protecting}.
	For apps, this means that credentials should not be available to other apps on the system.
	Finally, it also means that if the app is hosting a webpage in a WebView control, that the webpage cannot access credentials intended for the app, nor vice-versa.

While \prop{3} and \prop{1} help protect against many of the same attacks, leveraging both provides defense-in-depth.
This is especially important as both the current paper and past work~\cite{li2014emperor,silver2014password,stock2014protecting,oesch2020that} demonstrate that \prop{3} is poorly supported in most contexts.

\subsection{WebView}
\label{sec:background-webview}

WebView controls are UI widgets provided by the mobile operating system that serve as embeddable, minimalist browsers.
Apps use them to display web content directly within the app instead of having users click a link that opens the main mobile browser.
On Android, WebViews are added using the \texttt{WebView}\footnote{https://developer.android.com/guide/webapps/webview} class, whereas on iOS they are added using the \texttt{WKWebView}\footnote{https://developer.apple.com/documentation/webkit/wkwebview} class.
In both cases, the WebView control is implemented using WebKit.

To allow for integration between a hosting app and the content in the WebView, both iOS and Android allow the WebView control to be styled by the hosting app.
This styling is arbitrary and can even go so far as to make it impossible to distinguish between native widgets and content displayed in the WebView.
As such, the app can prevent the user from knowing they are interacting with content from a domain the app should not have access to, making phishing attacks trivial~\cite{luo2012touchjacking}.

Additionally, the hosting app can inject JavaScript into the content hosted in the WebView.
While this is intended to enable bidirectional communication between the app and the WebView content, in practice, it allows the app arbitrary control of the WebView content.
For security reasons, the Android developer documentation recommends that WebView only be used to show trusted, \emph{first-party} content.
However, prior work by Yang et al.~\cite{yang2019iframes} found that many popular apps---such as Google News, Facebook, and Uber---load third-party, untrusted content in a WebView and that 11K of the 17K most popular free apps on Google Play contained entry points to WebView loading APIs.

%% file: methodology.tex

\section{Evaluation Methodology}

There are three mobile autofill frameworks available on iOS and Android (the dominant mobile platforms).
On iOS, there is the app extensions framework and the Password AutoFill framework.
On Android, there is the autofill service.
Prior to the availability of the Android autofill service, many password managers used the Android accessibility (a11y) service to hack in support for autofill.
We chose not to include the a11y service in our evaluation for two reasons: first, it is not and designed as an autofill framework and thus does not serve as a meaningful point of comparison and second, it is now well-known that the accessibility service is ill-suited to be used for security purposes~\cite{jang2014a11y,lee2016study,fratantonio2017cloak,naseri2019accessileaks}.

As part of our evaluation, we identified three contexts in which autofill occurs on mobile devices:
\begin{enumerate}
	\item Webpages displayed in mobile browsers.
	\item Native UI elements (i.e., widgets provided by the OS) presented within mobile apps.
	\item Webpages displayed in WebView controls presented within mobile apps.
\end{enumerate}

In our evaluation, we consider all three of these contexts.
While the requirements to satisfy \prop{1} are the same for each of these contexts, they diverge for \prop{2} and \prop{3}.
As such, we used different tests in each context to explore the security of autofill.
These context-dependent tests along with their results are given in \S\ref{sec:browser}, \S\ref{sec:app}, and \S\ref{sec:webview}, respectively.

\subsection{Mobile Autofill Frameworks}

Below, we give a brief overview of each autofill framework we studied.

\input{appendix-systems}

\subsection{Testing Approach}

\begin{table}[t]
	\setuptable
	\renewcommand*{\prtline}{\hhline{-|---|--|}}
	
	\rowcolors{4}{gray!10}{}
	\begin{tabular}{l|lll|ll|}
		
		&
		\headrow{iOS Password AutoFill} & \headrow{iOS app extensions} & \headrow{Android autofill service} &
		\multicolumn{1}{c}{iOS} & \multicolumn{1}{c}{Android} \\ \hhline{~|---|--|}
		
		\multicolumn{1}{l|}{System} &
		\multicolumn{3}{c|}{Framework} &
		\multicolumn{2}{c|}{Version} \\ \hline
		
		1Password				
		&\checked 	&\checked	&\checked
		&7.4.7 		&7.4		\\
		
		Avast Passwords		
		&\checked 	&\checked	&\checked
		&1.15.4 	&1.6.4		\\
		
		Bitwarden				
		&\checked 	&\checked	&\checked
		&2.3.1 		&2.2.8		\\
		
		Dashlane				
		&\checked 	&\na		&\checked
		&6.2013.0 	&6.2006.3	\\ 
		
		Enpass				
		&\checked 	&\checked	&\checked
		&6.4.2 		&6.4.0		\\ \prtline
		
		iCloud Keychain 		
		&\checked 	&\na		&\na
		&13.3.1 	&---		\\
		
		Keepass2Android	 	
		&\na	 	&\na		&\checked
		&---		&1.07b-r0	\\ 
		
		Keeper				
		&\checked 	&\checked	&\checked
		&14.9.1 	&14.5.20	\\
		
		LastPass				
		&\checked 	&\checked	&\checked
		&4.8.0 		&4.11.4		\\ 
		
		Norton 				
		&\checked 	&\checked	&\checked
		&6.8.78 	&6.5.2		\\ \prtline
		
		RoboForm 				
		&\checked 	&\checked	&\checked
		&8.9.2 		&8.10.4		\\
		
		SafeInCloud 			
		&\checked 	&\na		&\checked
		&20.0.1 	&20.2.1		\\
		
		Smart Lock 			
		&\na	 	&\na		&\checked
		&--- 		&9.0		\\
		
		StrongBox				 			
		&\checked 	&\na		&\na
		&1.47.4 	&---		\\ \hline
		
	\end{tabular}
	
	\caption{Analyzed password managers on iOS and Android}
	\label{tab:systems}
\end{table}

Our evaluation of the three mobile autofill frameworks was primarily empirical in nature.
More specifically, we selected and evaluated 14 mobile password managers, each of which was implemented using one or more of the frameworks under test.
These managers were chosen as they are the most popular password managers implemented with the frameworks under test.
Table~\ref{tab:systems} summarizes these password managers, which frameworks they support, and which versions were tested.
Download counts for each tool are given in Appendix~\ref{appx:downloads}.

For each of these managers, we conducted a series of tests designed to evaluate how well the manager enforced \prop{1}--\prop{3} (see \S\ref{sec:secure-autofill}).
Within these evaluations, we paid attention to when the results were either the same and when they were different, helping us measure to what extent \prop{1}--\prop{3} were enforced by the framework, to what extent the frameworks left implementing these properties to the individual managers, and to what extent the frameworks limited the ability of the individual managers to achieve these properties.
In addition to this empirical evaluation, we also reviewed the documentation and APIs for each framework to try and contextualize and confirm our results.
We also reached out to developers of the password managers to understand the results, though only a few responded to our requests for information.

Testing in iOS was performed using an iPhone 7 running iOS 13.
Testing on Android was conducted using the Genymotion Android Emulator to simulate a Google Pixel 2 device running Android 9 (Pie).


%% file: appendix-systems.tex
\subsubsection{iOS App Extensions}

\begin{figure}
	\begin{subfigure}{.45\columnwidth}
		\fbox{\includegraphics[width=\textwidth]{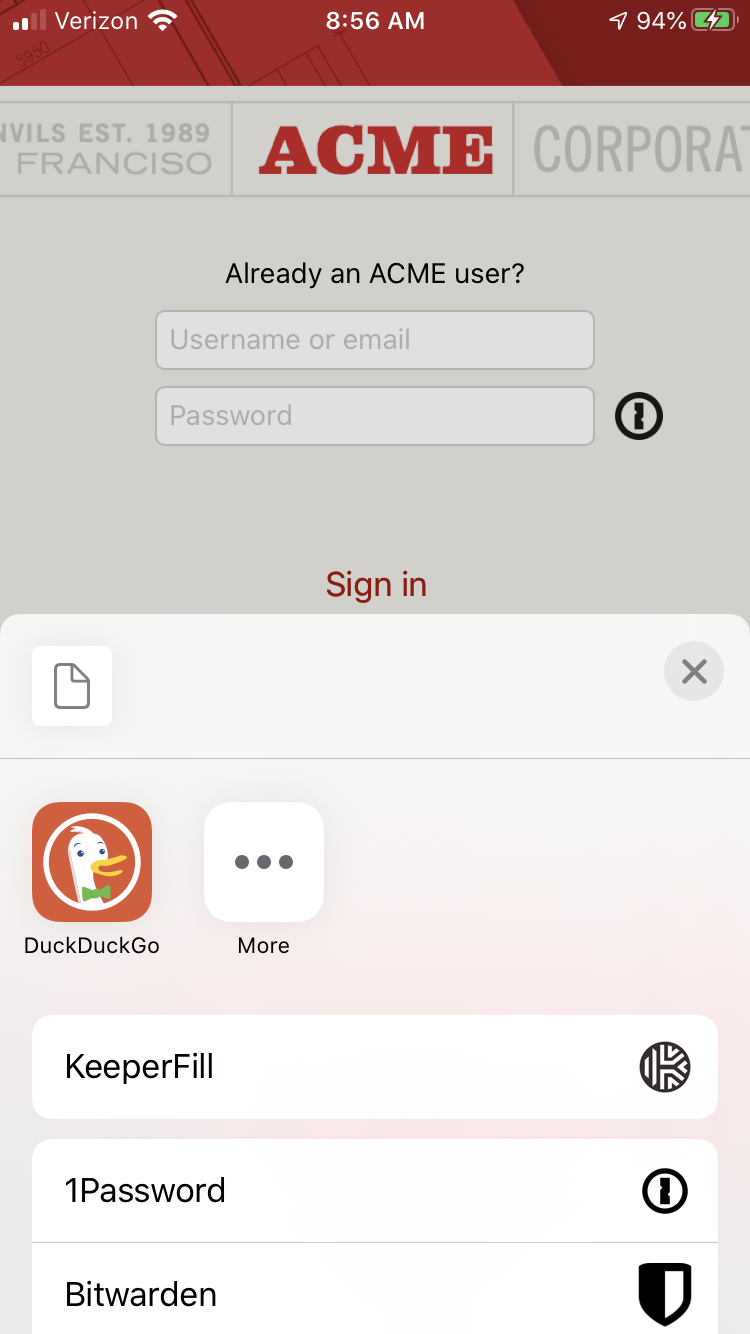}}
		\caption{Selecting app extension}
		\label{fig:selectextension}
	\end{subfigure}\hfil
	\begin{subfigure}{.45\columnwidth}
		\fbox{\includegraphics[width=\textwidth]{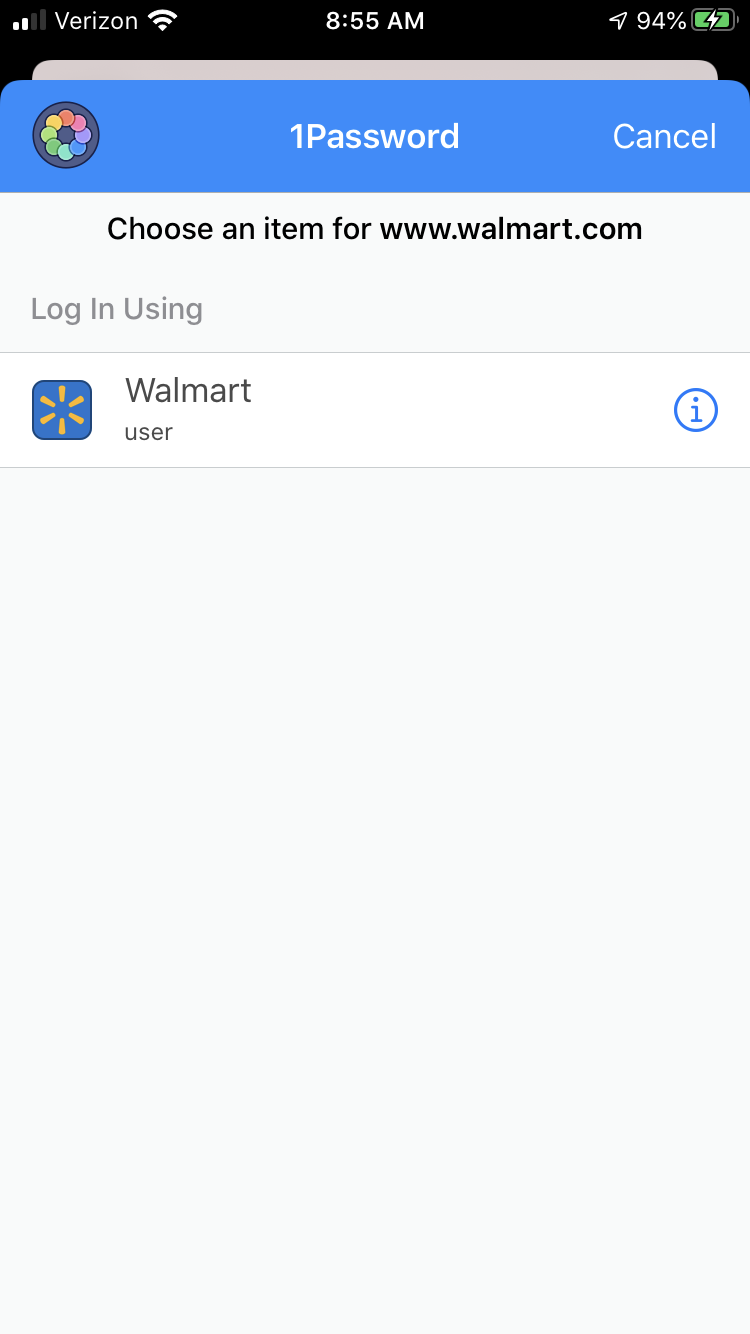}}
		\caption{Selecting password}
		\label{fig:selectcredential}
	\end{subfigure}
	
	\caption{iOS App Extensions UI}
	\label{fig:appextensionsui}
	\Description{Images demonstrating how users select an app extension and a password using the iOS App Extensions framework.}
\end{figure}

App extensions were introduced in iOS 8 (2014) and allow a host app to interact with another app (e.g., a password manager) using a predefined set of extension features.
For password autofill, this requires the password manager to implement the set of functions associated with the password management extension feature and for host apps to be updated to query this extension feature.
Note, the functionality provided by app extensions is minimal, enabling autofill, but not attempting to secure it.
For example, host apps are trusted to identify which credentials they should receive, with the framework doing nothing to check this mapping or verifying that the host app is not sending those credentials to a different domain.
Figure~\ref{fig:appextensionsui} shows the interface for app extensions, first requiring the user to select which app extension to use (see Figure~\ref{fig:selectextension}) and then selecting the credentials (see Figure~\ref{fig:selectcredential}).

While superseded by iOS Password AutoFill, we included iOS app extensions in our evaluation for three reasons: (1) it is still supported in iOS and remains functional in some password managers (e.g., 1Password, Keeper, LastPass) and host apps (e.g., Safari, Edge); (2) for older devices that cannot be updated to iOS 12, app extensions remain the preferred method for password autofill; (3) it provides a distinct approach to designing frameworks which provides a helpful point of comparison to the other two frameworks. 

\subsubsection{iOS Password AutoFill}

\begin{figure}
	\begin{subfigure}{.45\columnwidth}
		\fbox{\includegraphics[width=\textwidth]{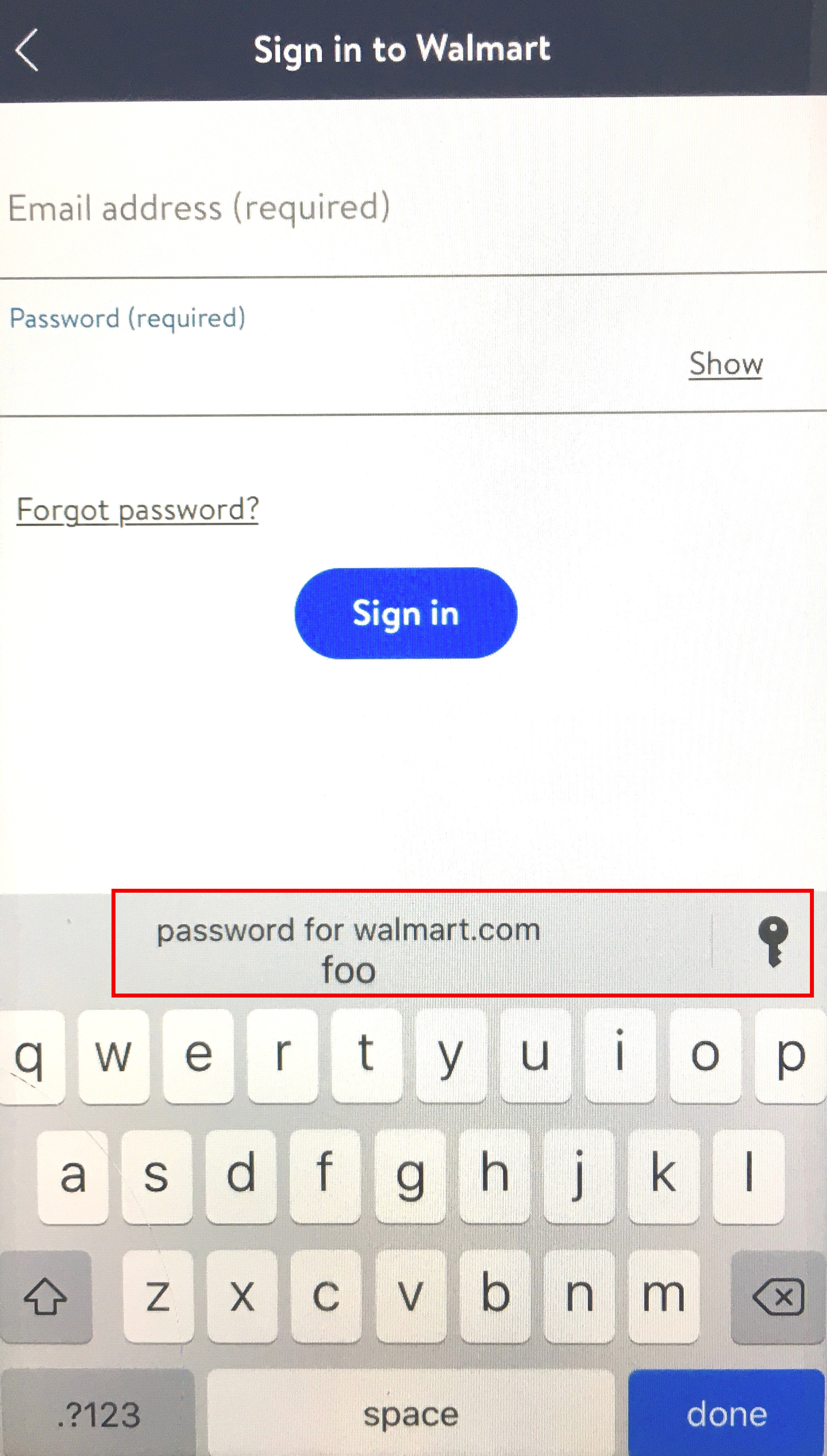}}
		\caption{Credential found}
		\label{fig:rec}
	\end{subfigure}\hfil
	\begin{subfigure}{.45\columnwidth}
		\fbox{\includegraphics[width=\textwidth]{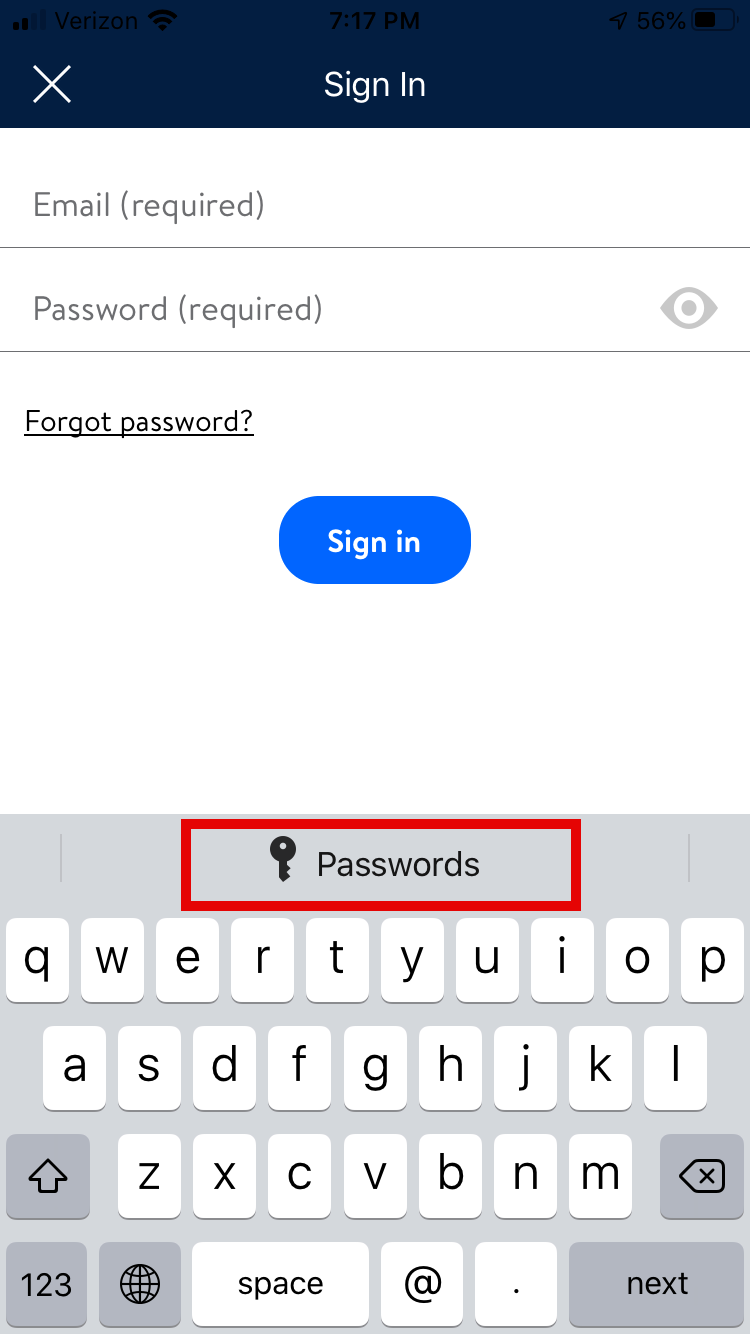}}
		\caption{No credential found}
		\label{fig:norec}
	\end{subfigure}
	
	\caption{iOS Password AutoFill UI}
	\label{fig:autofillui}
	\Description{AutoFill UI for iOS Password AutoFill.}
\end{figure}

The Password AutoFill framework was introduced in iOS 12 (2018) and takes a radically different approach to autofill.
Whereas app extensions provided minimal functionality, Password AutoFill controls the entire autofill, attempting to improve the usability and security of password autofill.
First, it handles the identification of login forms in both apps and websites, though the host app can help this process by annotating appropriate fields using the \texttt{textContentType} attribute.
Second, Password AutoFill ensures a secure mapping between an app and the domains that should have their credentials entered into that app.
That is done by having app developers include an Associated Domains Entitlement that indicates which domains are associated with the app; the domain operator is also required to include an \texttt{apple-app-site-association} file on their website indicating which apps can use credentials for that domain.
Third, Password AutoFill handles both the UI shown to users and the actual entering of credentials into the target app.
Figure~\ref{fig:autofillui} shows the interface for Password AutoFill, both when an associated domain can be found (see Figure~\ref{fig:rec}) and when not (see Figure~\ref{fig:norec}).

\subsubsection{Android Autofill Service}

\begin{figure}
	\begin{subfigure}{.45\columnwidth}
		\fbox{\includegraphics[width=\textwidth]{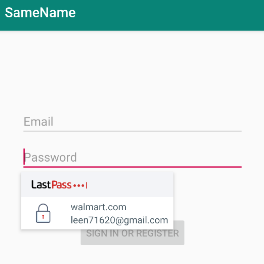}}
		\caption{LastPass}
		\label{fig:androidlastpass}
	\end{subfigure}\hfil
	\begin{subfigure}{.45\columnwidth}
		\fbox{\includegraphics[width=\textwidth]{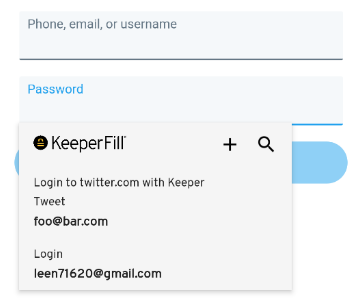}}
		\caption{Keeper}
		\label{fig:androidkeeper}
	\end{subfigure}
	
	\caption{Android Autofill Service UI}
	\label{fig:androidautofillui}
	\Description{Autofill service UI on Android.}
\end{figure}

In 2017, Android introduced the autofill service as part of API 26 (Android 8.0---Oreo).
This service falls between iOS's app extensions and Password AutoFill in terms of the features it supports.
Like Password AutoFill, it handles the identification and filling of login forms for apps and websites, with apps being able to help this process by annotating interfaces using the \texttt{android:autofillHints} attribute.
Unlike Password AutoFill, it does not control the credential selection UI (Figure~\ref{fig:androidautofillui} gives two examples of UIs provided by password managers on Android), nor does it enforce any app-to-domain credential mapping.

%% file: browser.tex

\section{Browser Autofill}
\label{sec:browser}

We begin our investigation by exploring autofill within mobile browsers.
As the security concerns for autofill on mobile browsers are the same as those on desktops, we base our methodology on the recent evaluation of desktop password managers conducted by Oesch and Ruoti~\cite{oesch2020that}.
We choose to use this methodology for two reasons---first, it is complete, covering all necessary aspects of browser autofill security, and second, it allows us to directly compare the performance of mobile password managers to desktop managers.
For each of the tests described by Oesch and Ruoti, we identify which of the three properties we synthesized (\prop{1}--\prop{3}) that the test relates to, removing extraneous tests that identify interesting edge cases but which cover features not needed to implement autofill securely.
We also considered if there were any additional tests needed to satisfy the three properties but found that the trimmed set of tests was sufficient to fully evaluate \prop{1}--\prop{3}.

All tests were performed using the default browser for each operating system: Safari (v13.3.1) for iOS and Chrome (v 81.0.4044) for Android.
We chose to focus on the default browser as they are likely the most widely used browser on each platform.
Additionally, the browsers are developed by the same company developing the autofill framework, allowing us to measure the security of the frameworks at their best.

In the remainder of this section, we describe the paired-down tests along with our results.
These results for each framework are summarized in Table~\ref{tab:browser}.
To allow for an easy comparison with desktop managers, Table~\ref{tab:browser} also provides the ratings for the most and least secure desktop managers from the prior work~\cite{oesch2020that}.
The full results for individual managers can be seen in Appendix~\ref{appx:results}.

\begin{table}[t]
	\setuptable
	\def\textast{\textsuperscript{\Large\textasteriskcentered}}
	
	\rowcolors{4}{gray!10}{}
	\adjustbox{width=\columnwidth}{
		\begin{tabular}{l|c|ccc|ccccc|}
			&
			\headrow{User interaction always required} &
			
			\headrow{Maps credentials to domains} &
			\headrow{Won't fill HTTPS$\rightarrow$HTTP} & \headrow{Won't fill HTTPS$\rightarrow$bad cert}  &
			
			\headrow{Fills password only on transmission} &
			\headrow{Won't fill different \texttt{action} (static)} & \headrow{Won't fill different \texttt{action} (dynamic)} & 
			\headrow{Won't fill different \texttt{method}} &
			\headrow{Won't fill cross-origin iframe}
			\\ \hhline{~|-|----|----|}
			
			Framework &
			\multicolumn{1}{c|}{\prop{1}} &
			\multicolumn{3}{c|}{\prop{2}} &
			\multicolumn{5}{c|}{\prop{3}}
			\\ \hline
			
			iOS Password AutoFill
			&\full
			&\full	&\none	&\none
			&\none	&\none	&\none	&\none	&\none \\

			iOS App Extensions
			&\full
			&\full	&\none	&\none
			&\none	&\none	&\none	&\none	&\full \\
			
			Android Autofill Service
			&\full	
			&\full	&\none	&\none
			&\none	&\none	&\none	&\none &\edit	\\ \hline
			
			Most secure desktop manager\textast
			&\full	
			&\full	&\prt	&\full
			&\none	&\prt	&\prt	&\prt	&\full \\
			
			Least secure desktop manager\textast
			&\none	
			&\full	&\full	&\none
			&\none	&\full	&\full	&\none	&\none \\ \hline
			
		\end{tabular}
	}
	
	\vspace{.5\baselineskip}
	\rowcolors{1}{}{}
	\begin{tabular}{ll}
		\full~~Secure behavior &
		\prt~~Partially secure behavior \\
		\none~~Insecure behavior &
		\edit~~Delegated to password manager
	\end{tabular}
	
	\vspace{.5\baselineskip}
	{\footnotesize \textast  Most and least secure desktop managers refer to the overall most and least secure managers---1Password and Firefox, respectively---from Oesch and Ruoti's work~\cite{oesch2020that}.}
	\vspace{.5\baselineskip}
	
	\caption{Autofill Security in Mobile Browsers}
	\label{tab:browser}
\end{table} 

\subsection{P1---User Interaction}

We tested whether this property was supported by constructing a login webpage and visiting it in the mobile browser, recording whether user interaction was required before credentials were filled.
We visited this webpage over HTTP, HTTPS, and using HTTPS with an invalid certificate to test whether this impacted user interaction requirements (as it does on some desktop browsers).

Our results show that all three frameworks correctly require user interaction before filling credentials.
This behavior is a marked improvement over desktop managers, where only 2 of the 12 managers tested by Oesch and Ruoti enforced user interaction.
This result shows the potential for autofill frameworks to enforce correct behavior across all password managers.

\finding{Within browsers, \prop{1} is enforced by all frameworks, a marked improvement over the situation on desktops.}

\subsection{P2---Credential-to-Domain Mapping}

To test credential mapping, we first registered credentials for different domains.
We then created testing webpages across multiple domains and checked that each domain only received approripate credentials.

We also tested if the credential mapping considered whether the webpage was authenticated using HTTPS.
To do this, we created test pages that were served over HTTP and over HTTPS with an invalid certificate, respectively, observing whether autofill would proceed or not.
Ideally, autofill would not be allowed in these cases, as such occurrences could represent a network attack being used to steal credentials.\footnote{The user can always override this behavior by manually copying and pasting credentials.}
We also allow for a rating of partially secure when autofill is allowed, but only after notifying the user of the potential danger.

We find that while all password managers map credentials to their appropriate domains, none of them check whether the webpage was served over a secure HTTPS connection, thus leaving open the possibility of network injection attacks.
Worse yet, the frameworks provide insufficient information to the password managers, preventing them from checking and enforcing this property themselves.
This leads to mobile managers being as bad as or worse than even the least secure desktop managers regarding \prop{2}.

\finding{Within browsers, \prop{2} is only partially enforced by the frameworks. The frameworks also prevent the managers from being able to enforce this property, causing mobile managers to be less secure than all desktop managers regarding \prop{2}.}

\subsection{P3---Protecting Filled Credentials}

The best way to achieve \prop{3} is to only fill credentials into the web request (where they are not accessible by JavaScript), not into the webpage, an approach outlined by Stock and Johns~\cite{stock2014protecting}.
Implementing this proposal is not possible on desktops as browsers do not allow extensions to modify web requests. 
In contrast, there should be no barrier for autofill frameworks to provide this functionality as the same entity maintains the framework and browser.
We test whether this proposal is implemented by trying to scrape credentials using JavaScript.

Without the above approach, it is not feasible to completely prevent malicious JavaScript from accessing the credentials enetered on a webpage.
Still, it is possible to limit the likelihood that those credentials will be accidentally sent to an unmapped domain.
First, the login form's \texttt{action} attribute should be checked to ensure that credentials will be submitted to the appropriate domain.
We test this by creating two webpages, one that has the form's action set to a different domain at page load (our static test) and one that sets the \texttt{action} field after page load, but before the credentials are autofilled (out dynamic test).

Second, the login form's \texttt{method} attribute should be checked to ensure that the credentials are not included in the URL (i.e., using a \texttt{GET} request), as this could potentially leak the credentials to another domain through the HTTP \texttt{Referrer} header.
We test this by creating a webpage with the method set to \texttt{GET} and observing whether autofill is allowed.
In both cases, we grade refusing to fill the credential as secure behavior, with a partially secure score being awarded if the credential is filled only after warning the user about potential dangers.

\begin{figure}[t]
	\centering
	\includegraphics[width=.8\columnwidth]{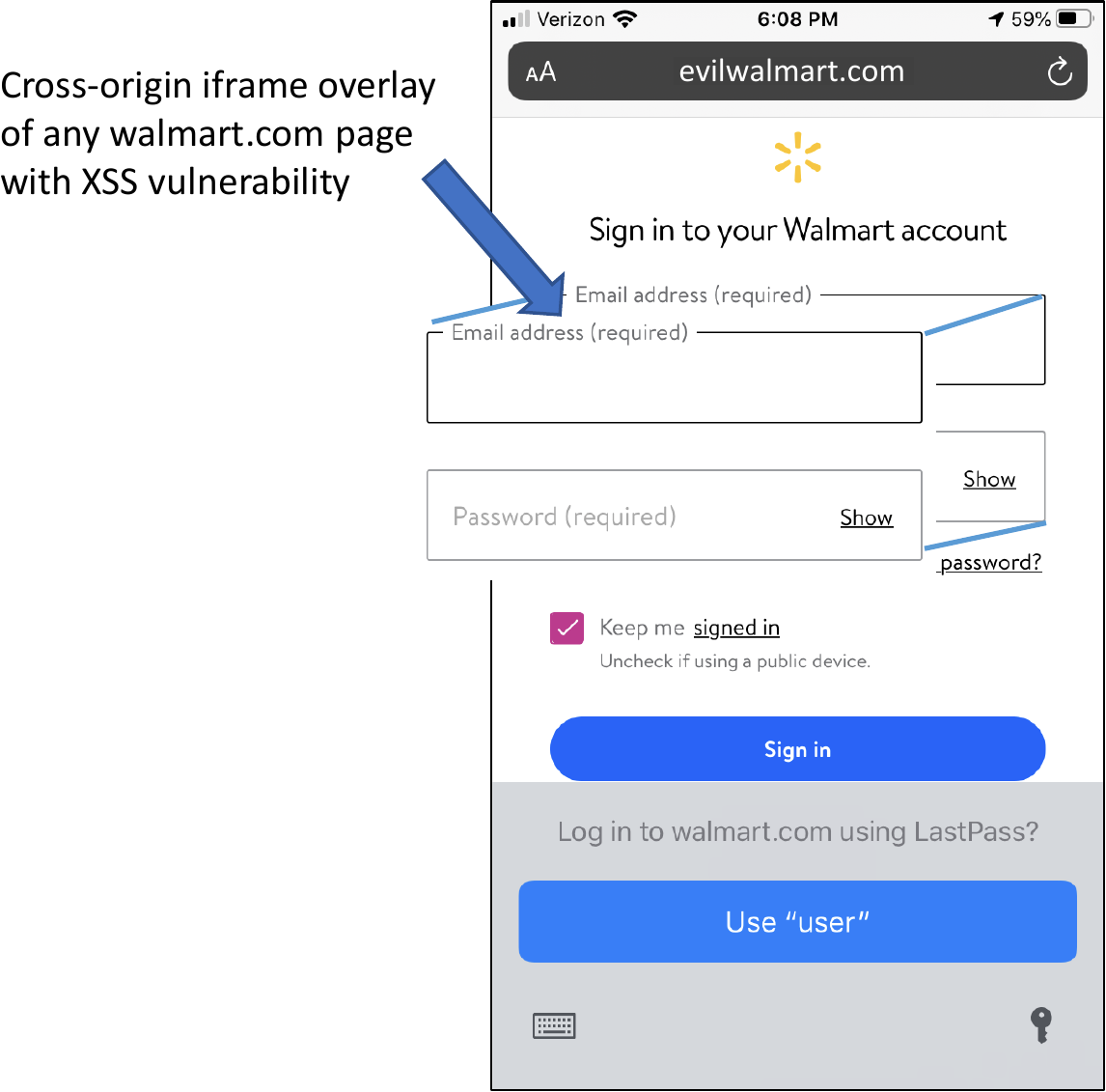}
	\caption{Example of a cross-origin iframe phishing attack}
	\label{fig:crossoriginattack}
	\Description{A cross-origin attack where an invisible iframe is used to trick users into autofilling credentials into a malicious domain instead of the intended domain.}
\end{figure}

Finally, autofill within cross-domain iframes should be disabled~\cite{silver2014password,stock2014protecting,oesch2020that}.
If a user visits a malicious or compromised webpage, the adversary can open a cross-origin iframe to any domain where they can inject JavaScript and steal credentials when those credentials are autofilled (see Figure~\ref{fig:crossoriginattack}).
In the worst case, if user interaction is not required, the credential theft will always succeed and unnoticed by the user.
Even if user interaction is required, the adversary is still able to effectively launch a phishing attack, which is likely to succeed as the appearance of an autofill dialog is supposed to indicate to the user that a phishing attack is not occurring (see \S\ref{sec:secure-autofill}).
We test this by creating a webpage with a cross-origin iframe and observe whether the iframe triggers autofill.

Our results show that none of the password managers adopt Stock and Johns' proposal.
Moreover, none of them check the \texttt{action} or \texttt{method} fields on the login form.
Similar to \prop{2}, the frameworks also fail to provide sufficient information to the managers to allow them to implement these features themselves.

Regarding cross-origin iframes, iOS Password AutoFill does not prevent cross-origin autofill and does not allow any of the managers to override this behavior.
In contrast, the older iOS app extensions properly prevent cross-origin autofill.
On Android, the autofill service does not prevent cross-origin autofill but does allow managers to override this behavior.
As with \prop{2}, the mobile frameworks perform as bad as or worse than even the least secure desktop manager regarding \prop{3}.

\finding{Within browsers, \prop{3} is not enforced, leaving filled credentials vulnerable to both theft and accidental leakage. The frameworks also prevent the managers from being able to enforce this property, causing mobile managers to be less secure than nearly all desktop managers regarding \prop{3}.}

%% file: app.tex

\section{App Autofill---Native UI Elements}
\label{sec:app}

We continue our evaluation by examining autofill for native UI elements within apps.
An overview of the results of our analysis is given in Table~\ref{tab:app}.

\begin{table}[t]
	\setuptable
	
	\rowcolors{4}{gray!10}{}
	\begin{tabular}{l|c|cc|cc|}
		&
		\headrow{User interaction always required} &
		
		\headrow{Secure app-to-domain mapping} &
		\headrow{Secure domain-to-app mapping} &
		
		\headrow{Prevents access from other apps} &
		\headrow{Prevents access from WebView}
		\\ \hhline{~|-|--|--|}
		
		Framework &
		\multicolumn{1}{c|}{\prop{1}} &
		\multicolumn{2}{c|}{\prop{2}} &
		\multicolumn{2}{c|}{\prop{3}}
		\\ \hline
		
		iOS Password AutoFill
		&\full
		&\full	&\full
		&\full	&\full \\
		
		iOS App Extensions
		&\full
		&\none	&\none
		&\full	&\full \\
		
		Android Autofill Service
		&\full
		&\edit	&\edit
		&\full	&\full \\ \hline
		
	\end{tabular}
	
	\vspace{.5\baselineskip}
	\centering
	\rowcolors{1}{}{}
	\begin{tabular}{ll}
		\full~~Secure behavior & \none~~Insecure behavior \\
		\multicolumn{2}{c}{\edit~~Delegated to password manager}
	\end{tabular}
	
	\caption{Autofill Security for Native UI Elements in Apps}
	\label{tab:app}
\end{table}

\subsection{P1---User Interaction}

We tested whether this property was supported by constructing a custom app and attempting to autofill it, recording whether user interaction was required before credentials were filled.
Our results show that all three frameworks correctly require user interaction before filling credentials.

\finding{Within native UI elements, \prop{1} is enforced by all frameworks.}

\subsection{P2---Credential-to-App Mapping}

There are two ways that credentials could be mapped to apps.
First, there could be a direct mapping between credentials and apps.
Second, there could be a mapping between Web domains and apps, leveraging the existing credential-to-domain mapping to provide an indirect credential-to-app mapping.
It is this second approach that is used by all mobile frameworks and managers.

Ideally, domain-to-app mappings should be bi-directional.
First, the app should identify which domains it is associated with, limiting the number of credentials it could access if the app became compromised (an application of the principle of least privilege).
Second, a domain should list which apps are allowed access to its associated credentials, preventing a malicious app from accessing arbitrary domains' credentials.
Only when both mappings agree should a credential for the given domain be suggested by the autofill framework.
Also, both mappings should be secured cryptographically---for example, by (1) code signing the app (including the file identifying the mapped domains), (2) using the fingerprint of the code signing certificate in the identification of apps on the domain's side, and (3) transmitting the domain's mappings using TLS.
Note, this bi-directional mapping does impede the use of single sign-on (SSO)---for example, Google's domain is unlikely to whitelist all the apps that use it for SSO---but this use case can be handled by hosting the SSO interface in a WebView (see \S\ref{sec:webview}).

To evaluate whether the frameworks satisfy these requirements, we first examined the documentation for each framework to understand what process they claimed to use and identify the mechanisms they used for app-to-domain mappings.
Our findings show that all three frameworks take a drastically different approach to \prop{2}, each requiring its own testing strategy.

\subsubsection{iOS Password AutoFill}
Apps indicate their associated domains by including an Associated Domains Entitlement file in their app package.
Since this file is part of the app package, its contents are signed as part of the code signing process required for all iOS apps.
Domains indicate the apps they are associated with by including an \texttt{apple-app-site-association} file at a specific URL on that domain.
This file indicates which apps are allowed to use credentials for that domain, with the appropriate code signing key for each app also being identified.
Credentials will only be autofilled if both mappings exist for a given domain.

To confirm that this functionality was working as intended, we created several testing apps and domains.
These apps included some with and without the appropriate mappings.
We also created look-alike apps with all the same information as a legitimate app but which were not signed with the correct code signing key.

Throughout all our testing, iOS Password AutoFill performed exactly as it should, providing a secure credential-to-app mapping and satisfying \prop{2}.

\subsubsection{iOS App Extensions}
The iOS app extension framework does not provide a mapping between apps and domains.
This behavior allows malicious apps to phish users' credentials, putting the onus on the user to detect that the malicious app should not receive the credentials suggested by the password manager, which, as previously discussed, runs counter to how autofill dialogs are supposed to work (see \S\ref{sec:secure-autofill}).
We verified this behavior by building an app that uses app extensions to autofill a login form and verifying that we could request credentials for arbitrary domains.

\subsubsection{Android Autofill Service}
The Android autofill service does not provide a mapping between apps and domains, instead leaving this functionality up to the individual managers to implement.
To analyze the mappings used by the 12 Android password managers we tested, we first inspected the autofill ceremony to see if domain-appropriate passwords were being suggested.
If they were, we used jadyx\footnote{\url{https://github.com/skylot/jadx}} to decompile the password manager's apk file and try to reverse engineer the credential mapping.
As part of this effort, we also used appmon\footnote{\url{https://github.com/dpnishant/appmon}} to intercept API calls from the password manager.

In our analysis, we identified four (non-exclusive) domain-to-app mappings used by the various managers: (a) a static list of app-to-domain mappings; (b) a custom heuristic that matches apps to domains based on the app's \texttt{applicationId}; (c) digital Asset Links (DAL) files hosted by domains at a specific URL are used to specify which apps---identified using their code signing key---should be mapped to that domain; and (d) relying on manual mappings provided by end-users.
Additional details about the implementation of each manager's mapping scheme is discussed in Appendix~\ref{appx:mapping}.

None of these mappings use a bidirectional app-to-domain mapping.
Only one mapping (DAL) requires domains to identify with which apps they are associated.
Only two managers check a cryptographic attestation of app identity, meaning that look-alike and side-loaded apps can impersonate legitimate apps and receive their associated credentials.
As such, our results demonstrate that while the autofill service's decision to delegating mappings could work in theory, in practice, it turns out to be a poor design decision.

\finding{For native UI elements, iOS AutoFill correctly provides \prop{2}. In stark contrast, iOS app extensions provide no credential-to-app mapping, allowing any app to request credentials for domains. The Android autofill service leaves \prop{2} up to managers to implement, but this turns out to be a bad decision with no manager correctly implementing such mappings.}

\subsection{P3---Protecting Filled Credentials}

Filled credentials should only be accessible to the app receiving those credentials, not to other apps or by webpages hosted in WebView controls within the filled app.
For all three frameworks, this property is satisfied by strong app segmentation guarantees provided by each framework's respective operating system.
Note, these protections can be side-stepped on Android using the accessibility service, but this is necessary to support individuals with disabilities, and as such, users need to remain careful about which apps they give permissions to control this service.

\finding{Within native UI elements, \prop{3} is enforced by the mobile operating system.}

%% file: webview.tex

\section{App Autofill---WebView Controls}
\label{sec:webview}

We end our investigation by considering autofill within WebView controls hosted inside apps.
Such functionality is critical to enable a range of use cases:

\begin{enumerate}
	\item \textbf{Supporting single sign-on (SSO).} 
	As described in \S\ref{sec:app}, autofill for native UI elements should only be allowed if the domain owner indicates the app is associated with the domain.
	As SSO providers are unlikely to whitelist every app that wants to use SSO, apps can instead use an embedded WebView control to display the SSO flow.
	
	\item \textbf{Thin wrapper apps.} Many mobile apps serve as little more than a thin wrapper around an existing website, with the app displaying a WebView control that displays the wrapped website.
	
	\item \textbf{Avoiding duplicate code.}
	Instead of having one authentication codebase for use on a website and one for the app associated with the website, app developers may instead choose to have authentication handled by the website using a WebView control.
\end{enumerate}

It is crucial to ensure that credentials are filled safely in each of these cases, especially for SSO credentials, whose theft would have an outsized effect.
An overview of our analysis of autofill security in WebView controls is given in Table~\ref{tab:webview}.

\begin{table}[t]
	\setuptable
	\newcommand{\highlight}[1]{#1}
	
	\rowcolors{4}{gray!10}{}
	\adjustbox{width=\columnwidth}{
		\begin{tabular}{l|c|ccc|cccccc|}
			&
			\headrow{\highlight{User interaction always required}} &
			
			\headrow{\highlight{Maps credentials to domains}} &
			\headrow{Won't fill HTTPS$\rightarrow$HTTP} & \headrow{Won't fill HTTPS$\rightarrow$bad cert}  &
			
			\headrow{\highlight{Prevents access from hosting app}} &
			\headrow{Fills password only on transmission} &
			\headrow{Won't fill different \texttt{action} (static)} & \headrow{Won't fill different \texttt{action} (dynamic)} & 
			\headrow{Won't fill different \texttt{method}} &
			\headrow{Won't fill cross-origin iframe}
			\\ \hhline{~|-|----|-----|}
			
			Framework &
			\multicolumn{1}{c|}{\prop{1}} &
			\multicolumn{3}{c|}{\prop{2}} &
			\multicolumn{6}{c|}{\prop{3}}
			\\ \hline
			
			iOS Password AutoFill
			&\full
			&\full	&\none	&\none
			&\none	&\none	&\none	&\none	&\none	&\none \\

			iOS App Extensions
			&\full
			&\edit	&\none	&\none
			&\none	&\none	&\none	&\none	&\none	&\full \\
			
			Android Autofill Service
			&\full	
			&\edit	&\none	&\none
			&\none	&\none	&\none	&\none	&\none &\edit	\\ \hline
			
		\end{tabular}
	}
	
	\vspace{.5\baselineskip}
	\rowcolors{1}{}{}
	\begin{tabular}{ll}
		\full~~Secure behavior & \none~~Insecure behavior \\
		\multicolumn{2}{c}{\edit~~Delegated to password manager}
	\end{tabular}
	
	
	\caption{Autofill Security for WebView Controls}
	\label{tab:webview}
\end{table} 

\subsection{P1---User Interaction}

We tested whether this property was supported by constructing a custom app with an embedded WebView control and attempting to autofill the WebView content, recording whether user interaction was required before credentials were filled.
Our results show that all three frameworks correctly require user interaction before filling credentials.

\finding{Within native UI elements, \prop{1} is enforced by all frameworks.}

\subsection{P2---Credential-to-Domain Mapping}
For WebView, credentials should be mapped based on the domain of the content displayed inside the WebView.
To test this, we created a test app mapped to one set of credentials, with the test app including a WebView with content from a domain associated with a different set of credentials.
We then tried to autofill a login form within the WebView and examined the set of credentials.
In addition to this test, we also replicate the two connection security tests used to evaluate mobile browser autofill.

Our results find that iOS Password AutoFill correctly maps credentials to the domain displayed in the WebView control.
iOS app extensions and the Android autofill service leave the mapping to individual managers, with only a minority using the correct mapping scheme.\footnote{iOS app extensions---1Password, Android autofill service---Dashlane, Keeper, Lockwise, SafeInCloud, and Smart Lock}
In contrast, most managers autofill credentials into the WebView based on the app's associated domains.\footnote{iOS app extensions---Keeper, Bitwarden, LastPass, and Enpass, Android autofill service---1Password, Bitwarden, Enpass, Keepass2Android, Keeper, LastPass, and RoboForm.}
This behavior leaves the app credentials vulnerable to phishing from compromised or malicious content displayed in the WebView, further described below.
The remaining managers refuse to autofill credentials into WebView controls, a significant limitation on utility.\footnote{iOS app extensions---Avast, Norton, and Roboform, Android autofill service---Avast.}

For the connection security tests, our results match the poor results for autofill within mobile browsers.\footnote{Android's WebView will not display HTTP content by default, though this can be overridden. When overridden autofill still works, so we graded this as insecure behavior.}

\subsubsection{Phishing Attack \#1: Using a Malicious Webpage}
In this phishing attack, a benign app uses a WebView to display content from a compromised domain---either because the attacker compromised a domain usually used by the app (e.g., XSS or supply chain attack) or because the attacker tricked the app into loading a domain of the attackers choosing.
Prior research has shown that such vulnerabilities are common in mobile apps~\cite{yang2019iframes}.
Once displayed, the malicious domain displays a (possibly hidden) login form, triggering the autofill ceremony and causing the password manager to suggest the user autofill the app's associated credentials (as opposed to the domain's associated credentials) into the WebView.
As the autofill dialog is intended to give confidence to users that it is safe to enter credentials (see \S\ref{sec:secure-autofill}), it is unlikely that they will even consider the possibility that a phishing attack is occurring.
Moreover, depending on the styling of the WebView and the content displayed therein, there may be no visual indication that the user is interacting with content not native to the app (see \S\ref{sec:background-webview}), eliminating nearly any chance that the user could detect a phishing attack.

To validate this attack, we developed a proof-of-concept app that displays content from a different domain in a WebView.
The displayed (malicious) domain presented a login form, styling it to look like part of the hosting app.
When the autofill framework suggested credentials, it was the app's, not the domain's associated credentials.
After clicking through the dialog, these credentials were sent to the malicious domain, successfully completing the phishing attack.
Based on our own experience with this proof-of-concept app and prior work~\cite{dhamija2006phishing,felt2011phishing,luo2012touchjacking,tuncay2020see}, we believe it should not be too difficult for adversaries to leverage this phishing attack in vulnerable managers.

\finding{Within WebView controls, only iOS Password AutoFill correctly maps credentials to the domain of the content in the WebView. The remaining frameworks leave this mapping up to the managers, with most managers using an incorrect mapping. This incorrect mapping leaves users vulnerable to a phishing attack where benign app credentials can be stolen by compromised content displayed in the WebView.}

\subsection{P3---Protecting Filled Credentials}
After filling credentials into the WebView, it should not be possible for the hosting app to extract those credentials.
Without this protection, it would be possible for a malicious app to host arbitrary login forms from various domains to steal those credentials.
To test this, we created a demo app mapped to one set of credentials, with the demo app including a WebView with content from a domain associated with a different set of credentials.
After autofilling credentials, we then attempted to use the hosting app to steal the credentials from the WebView.
In addition to this test, we also replicate the five \prop{3} tests used to evaluate mobile browser autofill.

We find that in each case, the WebView allowed the injection of malicious JavaScript into the WebView by the hosting app, with this JavaScript able to find and exfiltrate filled credentials.
This behavior leaves all of a user's credentials vulnerable to phishing by a malicious app.
For the remaining tests, our results match the poor results for autofill within mobile browsers.
Note, these problems could all be addressed by only filling credentials at the point they are transmitted to the server~\cite{stock2014protecting}, preventing the malicious JavaScript from accessing them on the webpage.

\subsubsection{Phishing Attack \#2: Using a Malicious App}
On both Android and iOS, the WebView control allows apps to inject JavaScript into any webpage displayed in a WebView.
A malicious app can use this feature to inject JavaScript that waits for credentials to be filled into the WebView and then exfiltrate them back to the malicious app.
After injecting the appropriate code, the malicious app triggers the autofill dialog by selecting the login elements hosted within the WebView, causing the autofill framework to suggest the user autofill the domain's associated credentials.
The adversary is also free to style the app, the WebView, and the content hosted in the WebView so that it is impossible for the user to tell that they are interacting with a WebView---for example, styling the WebView so that only a single text entry box (e.g., the username field) is shown, with no indication that the textbox it is not a native UI element.

As the autofill dialog is intended to give confidence to users that it is safe to enter credentials (see \S\ref{sec:secure-autofill}), it is unlikely that they will even consider the possibility that a phishing attack is occurring.
Still, if the adversary were to try to steal all the user's credentials immediately, this would likely be detected as the user would be bombarded with hundreds of autofill dialogs.
Instead, a careful adversary could stagger phishing attacks to instances when the user expects to authenticate within the app, stealing credentials over a long period.

We developed proof-of-concept apps for Android and iOS that implement this attack.
This app is styled to look like the Walmart app and phishes credentials when users attempt to log in.
Listing~\ref{lst:webview} shows the critical code used to implement the attack.
In both apps, we confirmed that our app was able to trigger autofill requests for arbitrary domains.
Additionally, based on our experience with this app and past research into phishing~\cite{dhamija2006phishing,felt2011phishing,luo2012touchjacking,tuncay2020see}, we believe it is unlikely that users would detect the attack.

\begin{figure*}
	\begin{subfigure}{\textwidth}
		\centering
		
		\begin{lstlisting}[language=swift,gobble=6]
			let controller = WKUserContentController()
			controller.add(self, name: "callbackHandler")
			
			func userContentController(_controller: WKUserContentController, didReceive message: WKScriptMessage) {
				if(message.name == "callbackHandler") {
					print("User credentials are \(message.body)")
				}
			}
		\end{lstlisting}
		\vspace{-.75\baselineskip}
		\caption{Malicious iOS app}
		\label{lst:webview-ios-app}
		\vspace{.5\baselineskip}
		
		\begin{lstlisting}[language=JavaScript,gobble=6]
			var username = document.getElementById("email").value;
			var password = document.getElementById("password").value;
			var credentials = `window.location.hostname:username:password`;
			window.webkit.messageHandlers.callbackHandler.postMessage(credentials);
		\end{lstlisting}
		\vspace{-.75\baselineskip}
		\caption{Injected JavaScript for iOS WebView}
		\label{lst:webview-ios-javascript}
		\vspace{.5\baselineskip}
		
		\begin{lstlisting}[language=Java,gobble=6]
			public class WebAppInterface {
				Context ctx;
				WebAppInterface(Context c) { ctx = c; }
				
				@JavascriptInterface
				public void stealCredential(String domain, String uname, String pword) {
					Toast.makeText(ctx, String.format(s:%s:%s", domain, uname, pword),Toast.LENGTH_SHORT).show();
				}
			}
		\end{lstlisting}
		\vspace{-.75\baselineskip}
		\caption{Malicious Android app}
		\label{lst:webview-android-app}
		\vspace{.5\baselineskip}
		
		\begin{lstlisting}[language=JavaScript,gobble=6]
			var uname = document.getElementById("email");
			var pword = document.getElementById("password");
			Android.stealCredential(window.location.hostname, uname.value, pword.value);
		\end{lstlisting}
		\vspace{-.75\baselineskip}
		\caption{Injected JavaScript for Android WebView}
		\label{lst:webview-android-javascript}
		\vspace{.5\baselineskip}	
		
	\end{subfigure}

	\captionof{lstlisting}{WebView Credential Exfiltration Scripts for iOS and Android}
	\label{lst:webview}
\end{figure*}

\finding{Within WebView, \prop{3} is not enforced, leaving filled credentials vulnerable to malicious apps and other Web vulnerabilities.}

%% file: rw.tex

\section{Related Work}
\label{subsec:relatedwork}

\textbf{Desktop Autofill Security:}
Silver et al.~\cite{silver2014password} studied the autofill feature of ten password managers.
They demonstrated that if a password manager autofilled passwords without requiring user interaction, it was possible to steal a user's credentials for all websites vulnerable to a network injection attack or had an XSS vulnerability on any page of the website.
They also showed that even if user interaction was required, if autofill was allowed inside an iframe, then the attacker could leverage clickjacking to achieve user interaction without users realizing they were approving the release of their credentials.
Stock and Johns~\cite{stock2014protecting} also studied autofill-related vulnerabilities in six browser-based password managers and had similar findings to Silver et al.
Li et al.~\cite{li2014emperor} studied five extension-based password managers and found logic and authorization errors, misunderstandings about the web security model, and CSRF/XSS attacks.

More recently, Oesch and Ruoti~\cite{oesch2020that} studied autofill in thirteen password managers, replicating and expanding previous work~\cite{silver2014password,stock2014protecting,li2014emperor}.
Their results showed that while modern password managers had addressed several key issues revealed by past studies, they remained vulnerable.

Within our work, we leverage the tests used by Oesch and Ruoti~\cite{oesch2020that} in our testing of autofill in mobile frameworks and WebView controls.
Using these tests allows us to compare our results to the results for desktop managers, demonstrating that even though these frameworks should perform better than desktop managers, they are worse than the least secure desktop managers.

\textbf{Mobile Autofill Security:}
Aonzo et al.~\cite{aonzo2018phishing} first demonstrated that the Android autofill service leaves app-to-domain mappings to individual managers.
They investigate five managers (Keeper, LastPass, 1Password, Dashlane, Smart Lock), finding that four of these could be tricked into suggesting to the user that they autofill credentials for a legitimate app into the malicious app.
Only Smart Lock avoided this pitfall by using cryptographic attestation.

Of the nine combinations of properties (\prop{1}--\prop{3}) and autofill contexts (mobile browsers, native UI elements in apps, WebView controls in apps), only one (\prop{2} for native UI elements) has been explored by this prior work.
Even in this one case, we expand on this prior work by (1) identifying the need for bidirectional app-to-domain mapping and (2) investigating how frameworks, as opposed to individual password managers, address this property.

%
%

\textbf{Feasibility of Phishing Attacks:}
Phishing attacks have been shown to be effective even against the most sophisticated users when visual deception, such as website redressing or overlays, is used~\cite{dhamija2006phishing}.
Felt and Wagner~\cite{felt2011phishing} also found that on mobile devices, users are often asked by legitimate apps and websites to autofill credentials after clicking a link.
As a result, users become conditioned to provide their credentials after clicking a link, making phishing attacks even easier. 
Luo et al.~\cite{luo2012touchjacking} also demonstrated several effective phishing attacks against WebView on iOS that utilized UI redressing and overlays to steal credentials. 
Most recently, Tuncay et al.~\cite{tuncay2020see} demonstrated that naming policies for Android apps allowed malicious apps to effectively masquerade as legitimate apps when requesting permissions from the user.

Several works have explored ways of protecting users from malicious apps that mimic the GUI of legitimate apps in an attempt to perform phishing or click-jacking attacks~\cite{bianchi2015app,fernandes2016android}.
Other prior work suggested content-based, and heuristic approaches for protecting mobile users from phishing websites~\cite{wu2014mobifish,shahriar2015mobile,goel2018mobile}.
To our knowledge, none of these solutions have been implemented, and phishing remains a problem.

This work is relevant to our paper as it shows that phishing attacks continue to work on mobile devices.
In particular, Tuncay et al.'s results showing that phishing attacks for permissions were successful on Android suggest that the credential phishing attacks described in our paper are also likely to be successful.

%
%

%% file: discussion.tex

\section{Discussion}\label{discussion}

Based on our findings, current mobile autofill frameworks are not achieving their potential.
However, we do not believe they should be abandoned but fixed to secure the autofill process properly.
Below we discuss (a) recommendations for addressing the WebView phishing attacks we identified, (b) guidelines for secure autofill framework implementations, and (c) areas requiring additional research.

\subsection{Addressing WebView Phishing Attacks}
One approach to addressing the WebView phishing attacks would be for the frameworks to adopt the proposal from Stock and Johns~\cite{stock2014protecting} to only fill credentials into Web requests, not the actual webpage.
Frameworks would implement this proposal by autofilling fake credentials into the webpage, replacing them with real credentials only when sent over the wire.
This behavior would prevent JavaScript, and by extension apps, from accessing the filled credentials.

Implementing this proposal on desktop environments is not currently possible as desktop browsers do not let extensions modify Web request contents.
In contrast, on mobile, the same vendor maintains the autofill framework, the mobile browser, and the WebView control.
This integration allows the vendor to implement Stock and Johns' proposal, and we strongly suggest they do so.
Note that implementing this feature would fully enforce \prop{3} for both WebView controls and the mobile browser.

Alternatively, autofill could be disabled for WebViews that have had JavaScript injected into them by the app.
Similarly, after autofill has occurred, injecting JavaScript could be disabled for the WebView until a new page is loaded (unloading the credentials).
These restrictions could be loosened for WebViews containing content from a domain if the domain is associated with the current app.

\subsection{Recommendations for Secure Autofill Frameworks} \label{sec:improvements}
Based on our results, we provide the following recommendations for autofill frameworks:

\begin{enumerate}[parsep=.5\baselineskip]
	\item \textbf{Require user interaction.}
	User interaction should always be required before credentials are autofilled. 
	This requirement ensures that users know when credentials are requested and entered, increasing the likelihood of detecting malicious activity.
	While this is far from a perfect defense, it does prevent silent credential harvesting and at least gives users a chance to detect a phishing attack.
	
	\item \textbf{Authenticate domains.}
	For both browsers and WebView controls, the domain's identity displayed in the WebView should be cryptographically verified using TLS.
	This verification will help prevent network injection attacks from being able to access filled credentials.
	
	\item \textbf{Provide a cryptographically verified bidirectional app-to-domain mapping.}
	Frameworks should require that apps identify the domains they are associated with, with this mapping file included in the code signing process.
	Domains should also be required to identify the apps that their credentials can be filled into, and this mapping should use the code signing key's fingerprint.
	Only when both these mappings agree should a given domain's credentials be autofilled into an app.
	
	iOS Password AutoFill already provides this property, and Android could provide this property by expanding their existing DAL-based app-to-domain pairing scheme, enforcing it at the framework level.
	While currently, adoption of DAL is limited---we analyzed 4,081 of the most popular paid apps and 9,345 of the most popular free apps on the Play Store and found that only 10\% of paid apps ($n=402$) and 20\% of free apps ($n=1879$) were whitelisted by a DAL file---we believe that adoption would rapidly increase if Google required such links for apps to be published or updated in the Google Play Store.
	
	\item \textbf{Thoroughly evaluate webpages.}
	Before filling in credentials, the framework should check all of the properties discussed in Section~\ref{sec:browser}.
	%
	For example, the framework should check the form to be autofilled, ensuring that the password is sent to the correct destination using HTTPS.
	Additionally, cross-domain autofill should be disabled.
	
	\item \textbf{Allow password managers to override autofill decisions.}
	When discussing our findings with password manager developers, we found that they were aware of many of the highlighted issues.
	However, these issues remain unresolved because the frameworks either prevent managers from gathering the information necessary to fix these problems (iOS app extensions and Android autofill service) or prevent them from changing the autofill process (iOS AutoFill framework).
	Manager security could be improved if the frameworks provided managers with more information about the autofill environment and process, such as providing information about the webpage where credentials are filled.
	Similarly, while the frameworks should provide safe defaults, managers could be allowed to override the frameworks' decisions to further restrict autofill as necessary.
	This behavior would return the locus of control to managers, allowing them to address an autofill framework's design flaws.
	
\end{enumerate}

\subsection{Future Research}
In addition to addressing the issues identified in this paper, we identify three areas of future research.

First, in this paper, we assume that users are likely to fall for the phishing attacks described in \S\ref{sec:webview}.
Based on how easily these phishing attacks can be obscured, the fact that the autofill dialog is supposed to indicate that phishing is not occurring, and the copious research establishing the ease of phishing users generally~\cite{dhamija2006phishing,felt2011phishing,luo2012touchjacking,tuncay2020see}, we believe our assumption is sound.
Still, future user studies could examine this attack, empirically confirming its feasibility.
Moreover, such research may also identify ways in which the autofill dialog could be improved to protect users from phishing attacks.

Second, we believe that research needs to be conducted to design a mechanism that allows a domain to indicate whichweb pages should receive autofilled credentials.
This functionality would prevent vulnerable webpages on the domain other than these login pages from stealing users' autofilled credentials, especially if password managers prevent autofill within same-origin iframes.
We believe this feature could be implemented similarly to how mappings work for iOS Password AutoFill or DAL files, having a single file on the website that lists acceptable URLs.
Still, research is needed to identify the feasibility and effectiveness of this proposal and the best way to implement it.

Third, research is needed on creating an autofill framework for desktop environments.
While browsers do provide a platform to deploy password manager extensions, they do not provide any password management-centric functionality---i.e., they do not assist with the detection of login forms nor facilitate autofilling credentials.
This lack of framework support causes a mixed level of security for password manager extensions.
Moreover, there is no OS-level autofill framework, making it nearly impossible for passwords managers to provide universal autofill for desktop applications.

%% file: conclusion.tex

\section{Conclusion}

Our analysis provides a mixed message regarding the effectiveness of mobile autofill frameworks.
On the positive side, all frameworks enforce user interaction before autofill (\prop{1}), significantly improving upon the situation on the desktop.
Additionally, iOS password autofill fully secures the autofill process for native UI elements in apps.

On the other hand, within mobile browsers, all frameworks failed to correctly check credential mapping (\prop{2}), and none adequately protected filled credentials (\prop{3}), in many cases being less secure than even the worst managers on desktop.
Moreover, the frameworks impeded the ability of managers to provide these properties themselves.
Thus, even with improvements for \prop{1}, this leaves mobile password managers less secure than their desktop counterparts.
These same issues cropped up for autofill within WebView controls in apps, with other issues leading to our identification of two phishing attacks enabled by the mobile autofill frameworks.
Critically, for both attacks, the password manager acts as a confused deputy, displaying the autofill dialog and suggesting that the user fills the credential being targeted by the attack, a dialog which in all other contexts indicates to the user that their credentials are not being phished (see \S\ref{sec:secure-autofill}).  

To us, these results represent an inflection point for autofill frameworks.
Either an immediate effort is needed to remedy the security flaws in these frameworks, or there is a need for these frameworks to be abandoned, allowing managers to secure the autofill process properly.
We strongly advocate for the prior approach, as if implemented correctly, these frameworks can ensure correct behavior across \textit{all} password managers.
Moreover, we also advocate for creating similar frameworks in browsers and desktop operating systems, allowing the benefits promised by frameworks to become universal.

%% file: appendix-downloads.tex

\section{Password Manager Download Statistics}
\label{appx:downloads}

\begin{figure}
	\centering
	\includegraphics[width=.9\columnwidth]{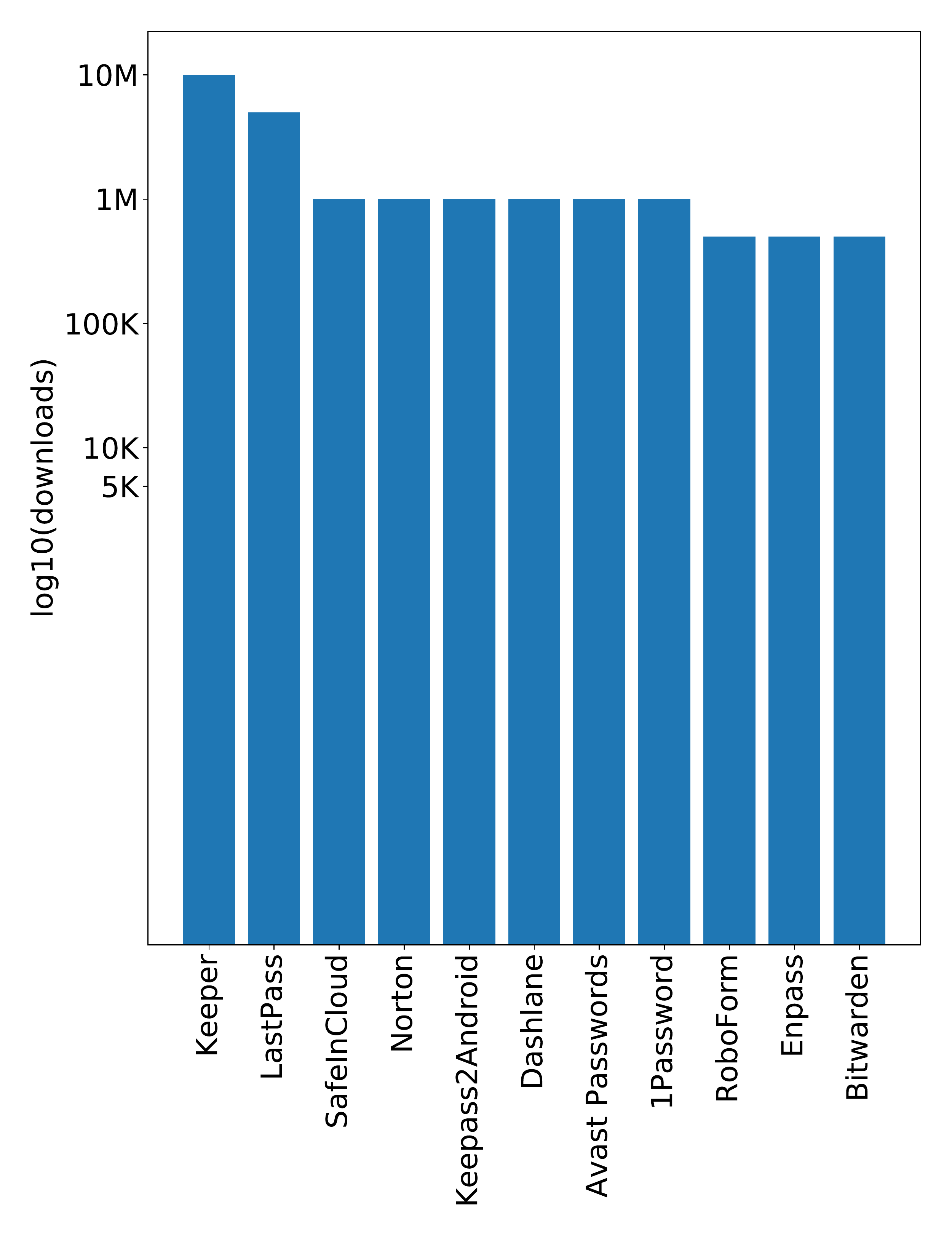}
	\captionof{figure}{Google Play Store Downloads from March 2020}
	\label{fig:androiddownloads}
	\Description{Download counts for password managers on the Google Play Store.}
\end{figure}

\begin{figure}
	\centering
	\includegraphics[width=.9\columnwidth]{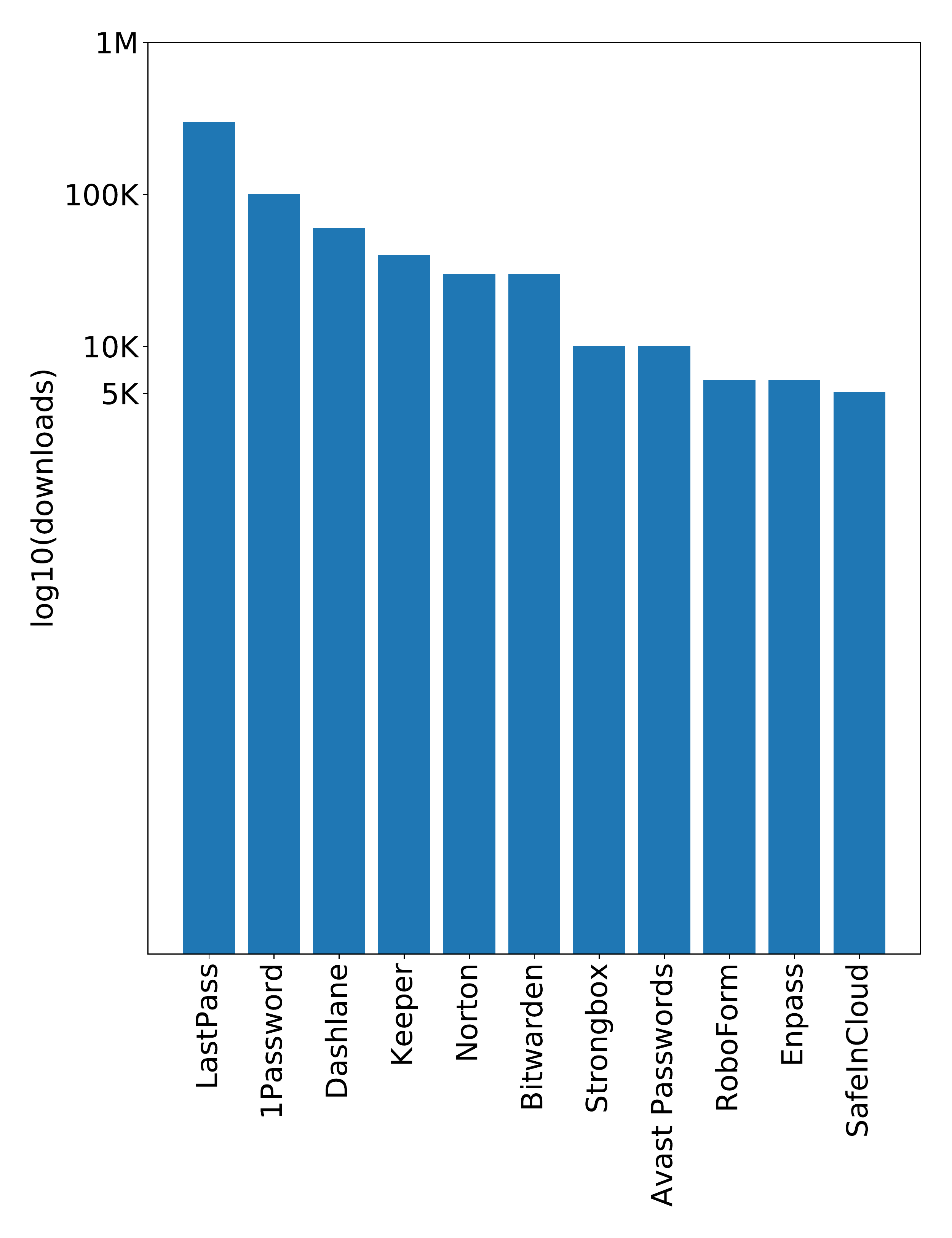}
	\captionof{figure}{iOS Download Estimates from April 2020}
	\label{fig:iosdownloads}
	\Description{Download counts for password managers on the Apple App Store.}
\end{figure}

This section presents the figures detailing app download statistics for the password managers we studied on iOS and Android.
On Android, we used the download count from the Google Play Store (see Figure~\ref{fig:androiddownloads}).
Because iOS does not provide detailed information about downloads from the App Store, we used estimates for April 2020 from SensorTower (see Figure~\ref{fig:iosdownloads}).\footnote{\url{https://sensortower.com/}} 

%% file: appendix-results.tex
\section{Password Manager Browser Evaluation Results}
\label{appx:results}

To test browser autofill for the mobile frameworks, we evaluated the security of fourteen different managers implemented with those frameworks.
This section gives detailed results for each manager using the evaluation criterion identified by Oesch and Ruoti~\cite{oesch2020that}.
Table~\ref{tab:autofill_in_browsers_iOS} gives the results of this evaluation for iOS and Table~\ref{tab:autofill_in_browsers_android} gives the results for Android.
Note that in Table~\ref{tab:autofill_in_browsers_iOS} there is only a single row for iOS autofill, as this framework completely handles the autofill experience for managers, obviating the need to report on the performance of individual managers (i.e., they are all the same).
In contrast, both iOS app extensions and the Android autofill service allow the managers to have limited control over the autofill process.

In addition to testing the various mobile managers, we also tested the password managers integrated into several mobile browsers.
While these managers do not use the system-wide autofill frameworks, they are a point of comparison for the managers implemented with the mobile frameworks.
These browser managers only work within their respective browsers and do not support generic app-based autofill.

\begin{table*}[p]
	\setuptable
	\renewcommand*{\prtline}{\hhline{~|-|---|--|---|----|---|--|}}
	
	\rowcolors{4}{gray!10}{}
	\adjustbox{height=1.875in}{
		\begin{tabular}{>{\cellcolor{white}}l|l|ccc|cc|ccc|cccc|ccc|cc|}
			\multicolumn{2}{l|}{} &
			\headrow{Interaction required for HTTPS} & \headrow{Interaction required for bad cert} & \headrow{Interaction required for HTTP} &
			\headrow{Won't fill same-origin iframe} & \headrow{Won't fill cross-origin iframe}  &
			\headrow{Won't fill different URL path} & \headrow{Won't fill HTTPS$\rightarrow$bad cert} & \headrow{Won't fill HTTPS$\rightarrow$HTTP}   & \headrow{Won't fill different \texttt{action} (static)} & \headrow{Won't fill different \texttt{action} (dynamic)} & \headrow{Won't fill different \texttt{method}} & \headrow{Won't autofill different \texttt{input} fields}  &
			\headrow{Won't fill \texttt{type="text"} field} & \headrow{Won't fill non-login form fields} & \headrow{Won't fill invisible password field}  &
			\headrow{Fills password on transmission} & \headrow{Obeys \texttt{autocomplete="off"}} 
			\\ \hhline{~~|---|--|---|----|---|--|}
			
			\multicolumn{1}{l}{}
			&\multicolumn{1}{l|}{System}
			&\multicolumn{3}{c|}{Interaction}
			&\multicolumn{2}{c|}{iframe}
			&\multicolumn{7}{c|}{Difference in fill form}
			&\multicolumn{3}{c|}{Fields}
			&\multicolumn{2}{c|}{Misc}
			\\ \hline
			
			\multicolumn{2}{l|}{Password AutoFill}
			&\full	&\full	&\full		
			&\none	&\none	
			&\none	&\none	&\none	&\none	&\none	&\none	&\none
			&\full	&\none	&\none
			&\none	&\none 	\\ \hline
			
			& 1Password
			&\full	&\full	&\full		
			&\full	&\full	
			&\none	&\none	&\prt	&\none	&\none	&\none	&\none
			&\full	&\none	&\none
			&\none	&\none 	\\
			
			& Avast
			&\full	&\full	&\full		
			&\full	&\full	
			&\none	&\none	&\none	&\none	&\none	&\none	&\full
			&\full	&\none	&\none
			&\none	&\none 	\\
			
			& Bitwarden
			&\full	&\full	&\full		
			&\full	&\full	
			&\none	&\none	&\none	&\none	&\none	&\none	&\none
			&\full	&\none	&\none
			&\none	&\none 	\\
			
			& Enpass
			&\full	&\full	&\full		
			&\none	&\full	
			&\none	&\none	&\none	&\none	&\none	&\none	&\none
			&\none	&\none	&\none
			&\none	&\none 	\\ \prtline
			
			& Keeper
			&\full	&\full	&\full		
			&\none	&\full	
			&\none	&\none	&\none	&\none	&\none	&\none	&\none
			&\full	&\none	&\none
			&\none	&\none 	\\
			
			& LastPass
			&\full	&\full	&\full		
			&\full	&\full	
			&\none	&\none	&\none	&\none	&\none	&\none	&\none
			&\full	&\none	&\none
			&\none	&\none 	\\ 
			
			& Norton
			&\full	&\full	&\full		
			&\full	&\full	
			&\none	&\none	&\none	&\none	&\none	&\none	&\none
			&\full	&\full	&\full
			&\none	&\none 	\\
			
			\headcol{-8}{App extensions}
			& RoboForm
			&\full	&\full	&\full		
			&\full	&\full	
			&\none	&\none	&\none	&\none	&\none	&\none	&\none
			&\full	&\none	&\none
			&\none	&\none 	\\ \hline
			
			& Chrome
			&\none	&\full	&\full		
			&\none	&\full	
			&\none	&\none	&\full	&\none	&\none	&\none	&\none
			&\full	&\none	&\none
			&\none	&\none 	\\
			
			& Firefox
			&\none	&\none	&\full		
			&\full	&\full	
			&\none	&\none	&\full	&\none	&\none	&\none	&\none
			&\full	&\none	&\none
			&\none	&\full 	\\
			
			\headcol{-3}{Browser\hspace{.3em}}
			& Edge
			&\full	&\full	&\full		
			&\full	&\full	
			&\none	&\none	&\none	&\none	&\none	&\none	&\none
			&\full	&\none	&\none
			&\none	&\none 	\\ \hline
			
		\end{tabular}
	}
	
	\vspace{.5\baselineskip}
	\rowcolors{1}{}{}
	\begin{tabular}{lll}
		\full~~Secure behavior &
		\prt~~Partially secure behavior &
		\none~~Insecure behavior
	\end{tabular}
	
	\caption{Autofill in mobile browsers on iOS}
	\label{tab:autofill_in_browsers_iOS}
\end{table*}

\begin{table*}[p]
	\setuptable
	\renewcommand*{\prtline}{\hhline{~|-|---|--|---|----|---|--|}}
	
	\rowcolors{4}{gray!10}{}
	\adjustbox{height=1.875in}{
		\begin{tabular}{>{\cellcolor{white}}l|l|ccc|cc|ccc|cccc|ccc|cc|}
			\multicolumn{2}{l|}{}  &
			\headrow{Interaction required for HTTPS} & \headrow{Interaction required for bad cert} & \headrow{Interaction required for HTTP}  &
			\headrow{Won't fill same-origin iframe} & \headrow{Won't fill cross-origin iframe}  &
			\headrow{Won't fill different URL path} & \headrow{Won't fill HTTPS$\rightarrow$bad cert} & \headrow{Won't fill HTTPS$\rightarrow$HTTP}   & \headrow{Won't fill different \texttt{action} (static)} & \headrow{Won't fill different \texttt{action} (dynamic)} & \headrow{Won't fill different \texttt{method}} & \headrow{Won't autofill different \texttt{input} fields}  &
			\headrow{Won't fill \texttt{type="text"} field} & \headrow{Won't fill non-login form fields} & \headrow{Won't fill invisible password field}  &
			\headrow{Fills password on transmission} & \headrow{Obeys \texttt{autocomplete="off"}} 
			\\ \hhline{~~|---|--|---|----|---|--|}
			
			\multicolumn{2}{l|}{System} &
			\multicolumn{3}{c|}{Interaction} &
			\multicolumn{2}{c|}{iframe} &
			\multicolumn{7}{c|}{Difference in fill form} &
			\multicolumn{3}{c|}{Fields} &
			\multicolumn{2}{c|}{Misc}
			\\ \hline

			& 1Password
			&\full	&\full	&\full		
			&\none	&\prt	
			&\none	&\none	&\none	&\none	&\none	&\none	&\none
			&\none	&\none	&\none
			&\none	&\none 	\\
			
			& Avast Passwords
			&\full	&\full	&\full		
			&\none	&\prt
			&\none	&\none	&\none	&\none	&\none	&\none	&\none
			&\full	&\none	&\none
			&\none	&\none	\\
			
			& Bitwarden
			&\full	&\full	&\full		
			&\none	&\prt	
			&\none	&\none	&\none	&\none	&\none	&\none	&\none
			&\none	&\none	&\none
			&\none	&\none	\\
			
			& Dashlane
			&\full	&\full	&\full		
			&\none	&\prt	
			&\none	&\none	&\none	&\none	&\none	&\none	&\none
			&\full	&\none	&\none
			&\none	&\none	\\ \prtline
			
			& Enpass
			&\full	&\full	&\full
			&\none	&\none	
			&\none	&\none	&\none	&\none	&\none	&\none	&\none
			&\none	&\full	&\none
			&\none	&\none \\
			
			& Keeper
			&\full	&\full	&\full		
			&\none	&\prt	
			&\none	&\none	&\none	&\none	&\none	&\none	&\none
			&\none	&\none	&\none
			&\none	&\none \\
			
			& LastPass
			&\full	&\full	&\full		
			&\none	&\prt	
			&\none	&\none	&\none	&\none	&\none	&\none	&\none
			&\none	&\full 	&\none
			&\none	&\none	\\
			
			
			& Norton
			&\full	&\full	&\full		
			&\none	&\prt
			&\none	&\none	&\none	&\none	&\none	&\none	&\none
			&\full	&\none	&\none
			&\none	&\none	\\
			
			& RoboForm
			&\full	&\full	&\full		
			&\none	&\prt	
			&\none	&\none	&\none	&\none	&\none	&\none	&\none
			&\full	&\none	&\none
			&\none	&\none	\\
			
			& SafeInCloud
			&\full	&\full	&\full		
			&\none	&\none
			&\none	&\none	&\none	&\none	&\none	&\none	&\none
			&\none	&\none	&\none
			&\none	&\none	\\
			
			\headcol{-12}{Autofill service}
			& Smart Lock
			&\none	&\none	&\none		
			&\none	&\none
			&\none	&\full	&\full	&\none	&\none	&\none	&\none
			&\full	&\none	&\none
			&\none	&\none	\\
			\hline
			
			& Chrome
			&\none	&\none	&\none		
			&\none	&\none
			&\none	&\full	&\full	&\none	&\none	&\none	&\none
			&\full	&\none	&\none
			&\none	&\none	\\
			
			& Firefox
			&\none	&\none	&\none		
			&\none	&\none
			&\none	&\none	&\none	&\none	&\none	&\none	&\none
			&\full	&\none	&\none
			&\none	&\none	\\
			
			\headcol{-3}{Browser\hspace{.3em}}
			& Opera
			&\none	&\none	&\none		
			&\none	&\none
			&\none	&\full	&\full	&\none	&\none	&\none	&\none
			&\full	&\none	&\none
			&\none	&\none	\\
			\hline
		\end{tabular}
	}
	
	\vspace{.5\baselineskip}
	\rowcolors{1}{}{}
	\begin{tabular}{lll}
		\full~~Secure behavior &
		\prt~~Partially secure behavior &
		\none~~Insecure behavior
	\end{tabular}
	
	\caption{Autofill in mobile browsers on Android}
	\label{tab:autofill_in_browsers_android}
\end{table*} 

%% file: appendix-mapping.tex

\section{Android Credential Mapping Details}\label{appx:mapping}

This section gives additional details on how password managers handled mapping apps and the domains associated with passwords stored in the password manager.

\textbf{1Password}, \textbf{Enpass}, \textbf{Keepass2Android}, and \textbf{RoboForm} require users to manually associate apps and domains.
Of these three, only RoboForm warns users of the danger that manual association can cause.

\textbf{Avast} maintains a SQLite database with two relevant tables.
The \texttt{domain\_info} table contains a list of 1,203 websites whose package names are a simple inversion of their website address.
For example, \texttt{facebook.com} is matched with \texttt{com.facebook}.
If the first two components of an app's package name are in this table, then the app is considered to match (e.g., \texttt{com.walmart.evil} is matched to \texttt{walmart.com}).
The second table, \texttt{alternate\_mapping} is a static mapping for apps that do not use a simple name inversion.
For example, this table maps \texttt{ign.com} to \texttt{com.mobile.ign}.

\textbf{Bitwarden} uses a simple heuristic that looks for substring matches between the domain and the components of the app's package name, though it does ignore components that are TLDs (e.g., \texttt{.com}, \texttt{.org}).
For example, \texttt{com.wal.evil} would match domains that contained \texttt{wal} or \texttt{evil}---for example, \texttt{walmart.com}.

\textbf{Dashlane} maintains a list of 285 mobile apps, their associated domains, and a SHA-512 hash of their signing certificates.
If an app is not on this list, Dashlane will not even offer to autofill it.
Unique to Dashlane, mapping behavior changes if the user turns off the autofill service and only enables the accessibility service.
In this case, instead of checking the list of allowed apps, Dashlane uses a simple heuristic that compares the components of the package name to domains looking for matches (ignoring common components such as \texttt{com} and \texttt{android}).
If a match is found, a warning is shown to the user informing them that they are autofilling an ``unverified app''.

\textbf{SafeInCloud} uses a simple heuristic that considers the first two components of the package name and matches those against domains.
For example, \texttt{com.walmart.evil} will match \texttt{walmart.com}.
While this does require malicious apps to use the same prefix as a legitimate app, we have confirmed that it is possible to upload such apps to the PlayStore.

\textbf{Smart Lock} maps apps by downloading the DAL files for the stored domains.
It did not allow a user to build an association between a website and an application.

\textbf{Keeper} uses a more complex heuristic to establish its credential mappings.
First, it will query the app store for information about the application.
If the app is found, Keeper will use the \texttt{app developer website} field as the domain for the app.
If the app is not found, the user will need to associate the app to a domain manually.
As first reported by Aonzo et al. ~\cite{aonzo2018phishing}, this heuristic is vulnerable to attackers who lie about the app developer website, something which is not verified when apps are uploaded.
We verified this by uploading an app to the Play store with the developer website set to \texttt{walmart.com} and checking that Keeper does indeed offer to fill Walmart's credentials into our app.
We note that Keeper does show a warning in this case.

\textbf{Lastpass} contains a SQLite database that maps apps and domains.
Additionally, like Lockwise and SafeInCloud, it uses a simple heuristic that considers the first two components of the package name and matches those against domains.
Unlike those two, if no match is found, the user is prompted to pick which domain should be matched with the app.
If the user does so, they are then asked if they want to share this mapping with other users.
If enough users share this mapping, it will be auto-suggested by LastPass to other users in the future (crowdsourced mappings).
LastPass does warn users when they first associate an app and a domain.

\textbf{Norton} includes a static file (\texttt{resources/assets/theirdpartyapp.properties}) mapping 131 package names to domains.
If an app is not on this list, Norton will not show an autofill dialog, not even to inform the user about the lack of a match.